\documentclass[preprint2,longabstract]{aastex}
\newcommand {\ia}{\'\i }

\def\xmm{{\em XMM--Newton}}
\def\chandra{{\em Chandra}}

\def\cm2{{cm$^{-2}$}}

\shorttitle{OTELO survey: II. Properties of X--ray emitters}
\shortauthors{Povi\'c et al.}

\begin{document}
 
\title{OTELO Survey: Deep BVRI broadband photometry of the Groth strip\\
II. Properties of X--ray Emitters}
\author{M. Povi\'c}
\affil{Instituto de Astrof\'isica de Canarias, 38205  La Laguna,  Spain}
\email{mpovic@iac.es} 
\author{M. S\'anchez-Portal}
\affil{Herschel Science Centre, ESAC/INSA, P.O. Box 78, 28691 Villanueva de la Ca\~nada, Madrid, Spain}
\email{miguel.sanchez@sciops.esa.int}
\author{A. M. P\'erez Garc\'ia}
\affil{Instituto de Astrof\'isica de Canarias, 38205  La Laguna,  Spain}
\author{A. Bongiovanni}
\affil{Instituto de Astrof\'isica de Canarias, 38205  La Laguna,  Spain}
\author{J. Cepa\altaffilmark{1}}
\affil{Departamento de Astrof{\ia}sica, Universidad de La Laguna, 38205  La Laguna,  Spain}
\author{J.A. Acosta-Pulido}
\affil{Instituto de Astrof\'isica de Canarias, 38205  La Laguna,  Spain}
\author{E. Alfaro}
\affil{Instituto de Astrof{\ia}sica de Andaluc{\ia}a-CSIC, Granada, Spain}
\author{H. Casta\~neda}
\affil{Instituto de Astrof\'isica de Canarias, 38205  La Laguna,  Spain}
\author{M. Fern\'andez Lorenzo}
\affil{Instituto de Astrof\'isica de Canarias, 38205  La Laguna,  Spain}
\author{J. Gallego}
\affil{   Departamento de Astrof{\ia}sica y CC. de la Atm\'osfera,
    Universidad Complutense de Madrid, Madrid, Spain }
\author{J. I. Gonz\'alez-Serrano}
\affil{Instituto de F{\ia}sica de Cantabria, CSIC-Universidad de Cantabria, Santander, Spain}
\author{J. J. Gonz\'alez}
\affil{Instituto de Astronom{\ia}a UNAM, M\'exico D.F, M\'exico}
\author{M. A. Lara-L\'opez}
\affil{Instituto de Astrof\'isica de Canarias, 38205  La Laguna,  Spain}

\altaffiltext{1}{Instituto de Astrof\'isica de Canarias, 38205  La Laguna,  Spain}

\begin{abstract}
The Groth field is one of the sky regions that will be targeted by the OTELO
   (OSIRIS Tunable Filter Emission Line Object) survey in the optical 820 nm
   and 920 nm atmospheric windows. This field has been observed by AEGIS (All--wavelength Extended Groth strip 
   International Survey) covering the full spectral range, from X--rays to radio waves. 
   \textit{Chandra} X--ray data  with total exposure time of 200ksec
   are analyzed and combined with optical broadband data of the Groth field
  in order to 
  study a set of structural parameters of the X--ray emitters and its relation with X--ray properties. 
  We processed the raw, public  X--ray data using the \textit{Chandra} Interactive Analysis of Observations 
 and determined and analyzed different structural parameters in order to produce a morphological classification of X--ray 
  sources. Finally, we analyzed the angular clustering of these sources using 2-point correlation functions.
   We present a catalog of 340 X--ray emitters with optical counterpart. We obtained the number counts and compared 
   them with AEGIS data. Objects have been classified by nuclear type
  using a diagnostic diagram relating X--ray-to-optical ratio (X/O) to hardness ratio (HR). Also, we combined 
  structural
  parameters with other X--ray and optical properties, and found for the first time an anticorrelation between 
  the X/O ratio and the
  Abraham concentration index which might suggest that early type galaxies have lower Eddington rates than those 
  of late type
  galaxies. A significant positive angular clustering was obtained from a preliminary analysis of 4 subsamples 
  of the X--ray sources
  catalog. The clustering signal of the opticaly extended counterparts is similar to that of strongly clustered 
  populations
  like red and very red galaxies, suggesting that the environment plays an important role in AGN phenomena.
\end{abstract}


\section{Introduction}

The OTELO project \citep{cepa05,cepa08} 
is a flux--limited survey of emission--line objects in large
and perfectly defined volumes of the Universe. The OTELO survey, more than one
magnitude deeper than other emission line surveys, includes a wide
variety of astronomical sources: star--forming galaxies, starburst galaxies, 
emission--line ellipticals, AGNs, QSOs, Ly$\alpha$ emitters, peculiar stars, etc. 
With these data, a number of scientific problems will be addressed, including: 
 star formation rates (SFR) and
metallicity evolution of galaxies; AGN and QSO evolution.  
Deep imaging programs such as several VLT surveys for GOODS \citep{dick04},
COSMOS \citep{capak07} and SXDS \citep{furusawa08}
fields will serve as preparatory work for OTELO.
Likewise, we have performed a deep BVRI broadband survey in one of the
fields of the Groth strip selected for the OTELO survey \citep{cepa08}. 

In the present work, data from this broad band survey is combined 
with other wavelength data for achieving several objectives of the OTELO project.
Specifically, the matching of these optical data with X--ray allows succesfully 
tackling the study of the AGN population as shown in \cite{barcons07}, \cite{geor06}, and \cite{steffen06}. 
Optical imaging surveys allow a morphological 
classification of all sufficiently resolved sources. This is
essential to reveal the nature of AGN host galaxies 
\citep[e.g.][]{hasan07,fiore01}, although 
the morphological classification becomes difficult at high redshift, due to the reduced resolution
of images.
Several works \citep{graham01,ferrarese00,gebhardt00} have shown that the AGNs are directly related with 
some host galaxy properties, 
particularly with their bulges. \citet{kauff03} have pointed out
that the AGNs in the local universe are hosted predominantly in bulge-dominated galaxies.
More recently, \cite{hasan07} using HST and {\it Chandra} data of the GOODS South field has found that
the most moderate luminosity AGNs hosts are bulge-dominated in the redshift range z\,$\sim$\,0.4--1.3.
A tight correlation between black hole mass and the bulge dispersion velocity has been well established
\citep{ferrarese00,gebhardt00}, as well as with the concentration of bulges \citep{graham01}.
 
On the other hand, deep X--ray extragalactic 
surveys are extremely useful tools to investigate the
time evolution of the population of active galactic nuclei (AGN) and to
shed light on their triggering mechanisms. 
The exceptional effectiveness at finding AGN arises largely because X--ray
selection has reduced absorption bias, minimal dilution by host-galaxy
starlight and allows efficient optical spectroscopic follow-ups
of high probability AGN candidates with faint optical counterparts. The {\it
Chandra} and {\it XMM-Newton} observatories have revolutionized this field of
research, generating the most sensitive X--ray surveys ever performed. In fact,
deep surveys produced by these observatories have increased the resolved
fraction of the cosmic X--ray background (CXRB) to about 90\% in the 0.5--2 keV
range. A large part of the detected X--ray sources are AGN ($\le$ 70\%), 
although X--ray emitters include a mix of different types of objects such as galaxy
clusters and stars. On the other hand, these powerful observatories make it
possible to access the hard (2--10 keV) range, allowing to directly probe the
AGN activity not contaminated  by star formation processes. Moreover,
hard X--ray surveys are capable to detect all but the most absorbed (Compton
thick) sources. Therefore, they provide the most complete and unbiased samples
of AGNs \citep{mush04,brandt05}. 
Number counts relations have now been determined \citep[e.g.][]{cappelluti07} although 
with some evidence for field-to-field variations. Such
variations are expected at some level owing to "cosmic statistics" associated with
large-scale structures and clustering features that have been detected in the X--ray sky 
\citep{barger03,yang03,gilli03,gilli05}.

In this paper we present the analysis of public, deep (200 ksec) \chandra/ACIS 
observations of three fields comprising the original Groth-Westphal  
strip (GWS), gathered from the Chandra Data Archive, and 
combined with optical BVRI data from our broadband survey 
carried out with the 4.2m William-Herschel Telescope (WHT) at La Palma \citep[see][for a detailed description]{cepa08}.
We have tried to implement an innovative approach to the optical study of X--ray emitters by investigating 
correlations between broadband optical and X--ray parameters, specifically the X--ray-to-optical ratio and hardness
ratios, and optical structural parameters that provide information about the host galaxy morphology and populations,
like asymmetry and concentration indexes and optical colors.

This paper is structured as follows: in Section \ref{observational_data}, we describe the observational data, including X--ray data processing and 
source detection, as well as optical broadband data and the selection of the sample of X--ray sources with optical
 counterparts. Section \ref{sec_analysis} reviews the X--ray properties of the galaxy sample, 
the procedures for computing optical structural parameters, as well as the morphological analysis carried out. The reliability of different structural parameters used in morphological classification is also discussed. 
Section \ref{sec_rel} presents the nuclear type classification, 
the relationship between X--ray and optical structural properties, 
and the comparison with results from the literature. The catalog of X--ray emitters with optical counterparts and their 
morphological classification is presented in the electronic edition. Finally, in Section \ref{clustering} we present
the results of the clustering analysis (based on the angular correlation function formalism) applied on 4 selected 
subsamples of the X--ray catalog.

\section{Observational data}
\label{observational_data}
\subsection{X--ray data}
\label{xray_data}
\subsubsection{Observations and data processing}
\label{xray_observations}
\chandra\  has observed three consecutive fields centered at the original 
HST Groth-Westphal strip (GWS) using the ACIS-I instrument.  All datasets have 
been gathered from the \chandra\ Data Archive 
(\texttt{http://asc.harvard.edu/cda/)} 
using the Chaser tool.  PI of all the retrieved \chandra\  observations is K. Nandra. 
Total exposure time in each field is about 200 ksec.  The total field size of
ACIS-I chips 0, 1, 2 \& 3 is 16.9'\,$\times$\,16.9' (ACIS-S chips 2 \& 3 were also used, but their
FOVs do not overlap with our optical data and therefore were not used). 
The \texttt{FAINT} and \texttt{VFAINT} telemetry modes have been used, with telemetry saturation rates of 170 and 67 events/s, respectively. Due to the lack of bright sources within 
the fields, we assume that there are no pile-up effects in any observation. 
Table \ref{tab_logx} summarizes the main characteristics of the \chandra\ observations.

\begin{table}[h]
\begin{center}
\caption{\chandra\ observations log for the three pointings of the GWS field. Col. (1): observation number; Col. (2,3): 
RA and DEC (J2000); Col. (4): Standard Data Processing (SDP) version; Col. (5): Observation format in the Time exposure mode; Col. (6): 
Net exposure time (Sum of Good Time Intervals, GTI) (ksec)
\label{tab_logx}}
\scriptsize{
\begin{tabular}{c c c c c c}    
\noalign{\smallskip}
\hline\hline
\noalign{\smallskip}
Obs&RA&DEC&SDP&Mode&GTI\\
(1)&(2)&(3)&(4)&(5)&(6)\\
\noalign{\smallskip}
\hline
\noalign{\smallskip}
5853&14:15:22.5&+52:08:26.4&7.6.7.1&VFAINT&42.4\\
5854&&&7.6.7.1&VFAINT&50.2\\
6222&&&7.6.7.1&VFAINT&35.2\\
6223&&&7.6.7.1&VFAINT&49\\
6366&&&7.6.7.1&VFAINT&15.96\\
7187&&&7.6.7.1&VFAINT&8\\
\noalign{\smallskip}
\hline
\hline
\noalign{\smallskip}
5851&14:16:24.5&+52:20:02.59&7.6.7.1&VFAINT&36\\
5852&&&7.6.4&VFAINT&11.95\\
6220&&&7.6.7.1&VFAINT&36\\
6221&&&7.6.7.1&VFAINT&3.8\\
6391&&&7.6.7.1&VFAINT&9.4\\
7169&&&7.6.4.1&VFAINT&18.2\\
7181&&&7.6.7.1&VFAINT&8\\
7188&&&7.6.4.1&VFAINT&4.2\\
7236&&&7.6.4&VFAINT&18.8\\
7237&&&7.6.4.1&VFAINT&17.2\\
7238&&&7.6.4&VFAINT&9.6\\
7239&&&7.6.4.1&VFAINT&16.2\\
\noalign{\smallskip}
\hline
\hline
\noalign{\smallskip}
3305&14:17:43.0&+52:28:25.2&6.8.0&FAINT&27.6\\
4357&&&6.9.0&FAINT&79.8\\
4365&&&6.9.0&FAINT&58.2\\
\noalign{\smallskip}
\hline
\hline
\end{tabular}
}
\normalsize
\rm
\end{center}
\end{table}

We processed the data Using the \chandra\  Interactive Analysis of 
Observations (CIAO) (\texttt{http://cxc.harvard.edu/ciao/}), v4.0 
and Calibration Data Base (CALDB) v3.4.0. 
Standard reduction procedures have been applied, as described in 
the CIAO Science Threads to produce new level 2 event files. The output level 2 event files have been 
restricted to the  0.5--8\,keV range to avoid high background spectral regions.
Furthermore, we analyzed the source-free light curves in order to define additional good time intervals (GTI). 
To this end, we applied the CIAO detection program $\texttt{celldetect}$ 
with a threshold S/N\,=\,3.  The detected sources have been removed upon creation of the time-binned light curves, 
that have been then used for the computation of the GTIs. A $\pm$3$\sigma$ rejection criterion has been applied to remove high background intervals. After filtering for the additional GTIs, the level 2 event files belonging to the same target were co-added. To this end, we first improved the absolute astrometry (the overall 90\% uncertainty circle of 
Chandra X--ray absolute position has a radius of 0.6 arcsec) applying corrections for translation, scale and rotation to the world coordinate system (WCS) of each data file, by comparing 
two sets of source lists from the same sky region. Average positional accuracy improvement is 6.3\% with respect to the original astrometry provided by the spacecraft attitude files. 
Once we have improved the astrometry, the event 
files have been merged.

The output merged files have been filtered to create event files for several 
energy bands: full (0.5--7\,keV), soft (0.5--2\,keV),  
hard (2--7\,keV) hard2 (2--4.5\,keV) and vhard (4--7\,keV). Unbinned effective exposure maps at a single energy, representative 
of each band, were created at 2.5\,keV (full), 1\,keV (soft), 4\,keV (hard), 
3\,keV (hard2) and 5.5\,keV (vhard). Resulting maps have units of 
time $\times$ effective area, i.e.  s\,cm$^2$\,counts\,photon$^{-1}$.

\subsubsection{Source detection}
\label{xray_detection}
We applied the CIAO \texttt{wavdetect} Mexican-Hat wavelet source 
detection\footnote{The fact that the PSF of X--ray detectors often has Gaussian-like shape motivates our use of the Marr wavelet, 
or ``Mexican Hat'' function.} program to all bands (full, soft, hard, hard2 and vhard) 
for the three fields. Several wavelet scales have been 
applied: 1, $\sqrt{2}$, 2, 2$\sqrt{2}$, 4, 4$\sqrt{2}$, 8, 8$\sqrt{2}$ and 16 pixels. Small scales are well suited to the detection of small sources while larger scales are appropriate for more extended sources. Due to the ``blank field'' nature of the GWS, 
we do not expect to detect sources at a scale larger than 16 pixels 
($\sim$ 8\,arcsec). For each applied scale,  $\texttt{wtransform}$ produces a complete set of detections. We set a significance threshold of  2\,$\times$\,10$^{-7}$. Since this \texttt{wavdetect} input parameter 
measures the number of spurious events per pixel, we expect $\approx$\,0.2 fake detections per detector of 
1024\,$\times$\,1024 pixels i.e. about 12 spurious sources in the entire catalog (4 detectors $\times$ 5 
bands $\times$ 3 fields).
The output \texttt{wavdetect} detection files are \textsc{fits} tables containing the position of sources 
(RA \& DEC), count rates in photons\,s$^{-1}$\,cm$^{-2}$ and ancillary information. The final X--ray source 
catalog has been obtained by cross--matching detections in all five bands in order to increase the reliability of detections. 
Best match is given by a minimum 
separation of sources around a great circle. We set a maximum search distance of 2\,arcsec 
(see section \ref{catalog}). We computed the hardness ratios as follows:

\begin{equation}
HR\left(\Delta_1E,\Delta_2E\right) = \frac{CR\left(\Delta_1E\right) - CR\left(\Delta_2E\right)}{CR\left(\Delta_1E\right) 
+ CR\left(\Delta_2E\right)}
\end{equation}

\noindent where $\Delta_1E$ y $\Delta_2E$ are different energy bands and $CR\left(\Delta_nE\right)$ is the
count rate in a given energy band. Four hardness ratios have been set in this way: HR$_1$\,$\equiv$ HR(hard, soft), 
HR$_1'$\,$\equiv$ HR(hard2, soft),
HR$_2$\,$\equiv$ HR(very hard, hard) and HR$_2'$\,$\equiv$ HR(very hard, hard2).\\
We also computed the energetic fluxes from count rates. To this end, mean photon energies in each band have 
been calculated assuming a power law spectrum with $\Gamma$\,=\,1.5 (a reasonable assumption after inspecting the hardness ratios diagram and
diagnosis grid depicted in Figure \ref{hr_grids}), redshift z\,=\,0.5, galactic absorption 
n$_H$\,=\,1.3\,$\times$\,10$^{20}$\,cm$^{-2}$ and no intrinsic absorption. We applied the absorption 
coefficients provided by the \texttt{TRANNM} function within the package PIMMS v3.9b (Mukai, 1993).
The output catalog has 639 unique X--ray emitters (after removing 20 common sources in the field overlapping area). 
From them, 76 objects are located outside the field covered by optical data. From the total sources,
46\% (297)  have been detected with a significance $\geq$ 4$\sigma$,  21\% (132) above the  
3$\sigma$ significance level and 13\% (81) down to 2.5$\sigma$ level.\\

The number of sources detected in each band are 490 (full), 465 (soft), 284 (hard), 266 (hard2) and 121 (vhard), 
with median fluxes of 1.81\,$\times$\,10$^{-15}$, 5.57\,$\times$\,10$^{-16}$, 2.34\,$\times$\,10$^{-15}$, 1.58\,
$\times$\,10$^{-15}$ and 3.75\,$\times$\,10$^{-15}$\,erg\,cm$^{-2}$\,s$^{-1}$, and with limiting fluxes at 3$\sigma$ 
level of 4.8\,$\times$\,10$^{-16}$, 1.1\,$\times$\,10$^{-16}$, 2.8\,$\times$\,10$^{-16}$, 1.3\,$\times$\,10$^{-16}$ 
and 7.3\,$\times$\,10$^{-16}$\,erg\,cm$^{-2}$\,s$^{-1}$, respectively. Median errors in fluxes (for objects above the 
3$\sigma$ significance level) are 8\% in full band, 
10\% in soft band, 12\% in hard band, 14\% in hard2 band and  22\% in vhard band.
There are 16 sources detected only in hard band, 18 in hard2 band and 3 sources are only observed in vhard band. 
We find an overabundance of soft X--ray sources in our final catalog. 
There are 103 sources detected only in this band. 
This effect can be at least partially explained considering the decrease of telescope\,+\,ACIS effective area 
from some 600 cm$^{2}$ in the 1--2\,keV range (including the 0.5--2\,keV band defined as soft), 
until approximately 200\,cm$^{2}$ at $\sim$\,6\,keV. On the other hand, the total background between 0.5 and 7\,keV 
is around 4 times higher than in the 0.5--2\,keV range. The combination of these two factors justifies 
the high \chandra/ACIS source detection efficiency in the soft X--ray band and the mentioned overabundance.
The first version of this catalog was presented in \cite{Sanchez07}.

\subsection{Comparison with AEGIS data}
\label{AEGIS_data}

\begin{figure}[ht!]
\centering
\includegraphics[width=7.5cm,angle=0]{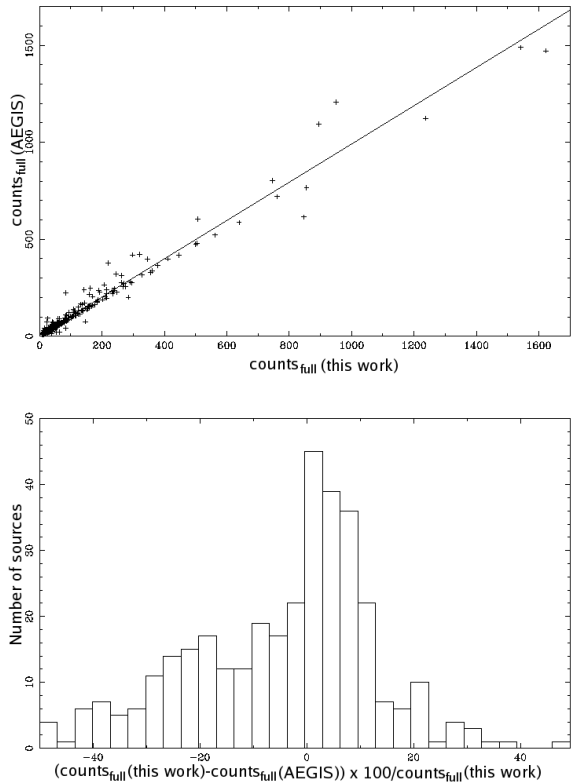}
\caption[ ]{Upper pannel: full band net counts from this work vs AEGIS net counts. The solid line represents the best linear fit, with a slope of 0.99 and zero point 5.23 counts. Lower pannel: histogram of relative deviations of our full band count data with respect to AEGIS. Median deviation is lower than 2\%.
\label{aegis_comparison}}
\end{figure}

The All-wavelength Extended Groth strip International Survey (AEGIS) collaboration has recently
made public their data sets, including a catalog of  X--ray sources that comprises the observations used
within this paper. The complete AEGIS X--ray catalog \citep{laird08} consists of 1325 sources; from
these, 471 are located within the area covered by our data (fields EGS-6, 7 and 8 according to their nomenclature).
From our 429 sources detected above 3$\sigma$ significance level, 416 coincide with AEGIS sources using a search
radius of 2\,arcsec (388 sources if only unique matches are considered).
Figure \ref{aegis_comparison} shows a good agreement between the source counts used in this work and those 
provided by the AEGIS team: median relative deviations of our data with respect to AEGIS are 2\% (full band), 0.7\% (soft),
5.5\% (hard) and 7.8\% (vhard). 

\subsection{Optical data}
\label{optical_data}

We observed three pointings in the direction of the GWS, 
each of them using the B, V, R, and I broadband filters. The total
area covered is 0.18 square degrees. The observations have been 
carried out using the Prime Focus Imaging Platform (PFIP) at the
4.2m William-Herschel Telescope (WHT) 
in the Roque de los Muchachos Observatory (La Palma, Canary Islands). 
This camera has two CCD detectors (2k\,$\times$\,4k), with a total field of
view of  16'\,$\times$\,16' and a  pixel size of 0.237 arcsec. 
Several exposures have been performed at each pointing, of 600, 800, 900, or 1000 sec, depending on the filter, with a dithering of 15\,arcsec between consecutive exposures 
in order to allow eliminating cosmic rays and to fill the gap
between detectors. Standard reduction procedures have been applied. 
Resulting limiting magnitudes are 25, 25, 24.5 and 23.5 in B, V, R and I band, respectively.
A detailed description of data reduction and optical source detection can
be found in \citet{cepa08}. The optical catalog used in this work 
contains data for $\sim$\,44000 objects. 

\subsection{The catalog of optical counterparts}
\label{catalog}
\begin{figure}[!ht]
\centering
\includegraphics[width=7.5cm,angle=0]{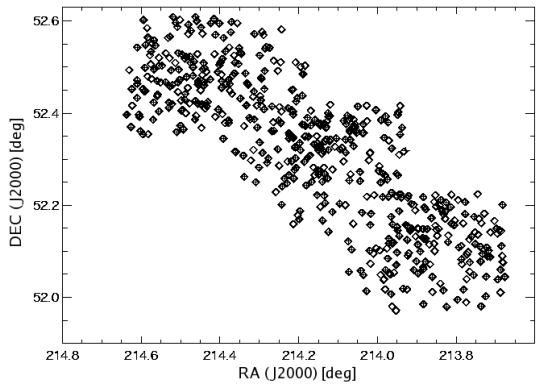}
\protect\caption[ ]{X--ray sources (diamonds) and their optical counterparts (filled diamonds) from the cross--match of X--ray and optical catalogs \label{xray_opt_counterparts}}
\end{figure}

\begin{figure}[!ht]
\centering
\includegraphics[width=7.5cm,angle=0]{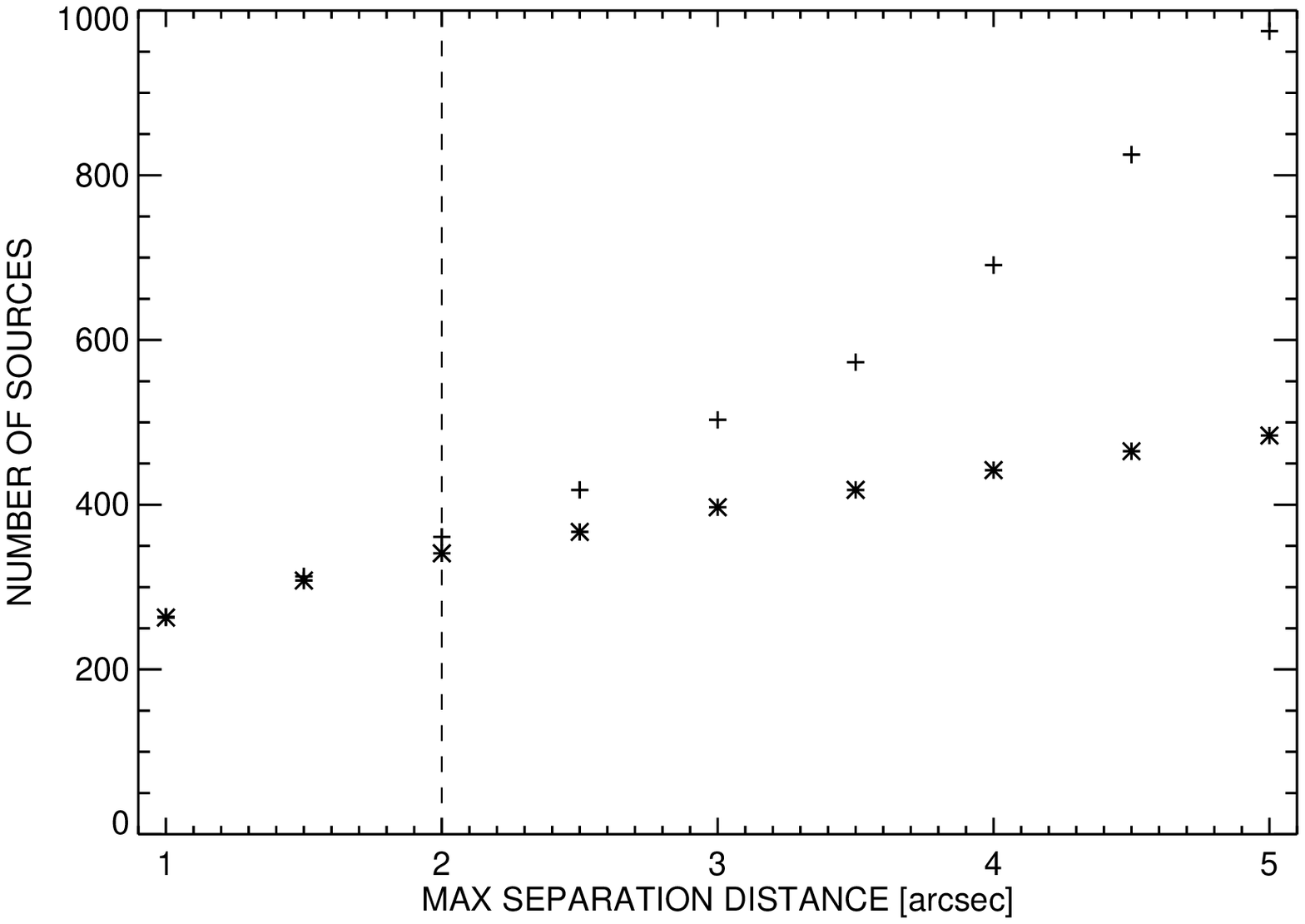}
\protect\caption[ ]{Number of optical counterparts as a function of the maximum search radius used for cross--matching. The number of closest sources is represented with stars. Crosses represent the total number of sources found within the search radius. The dashed line marks the adopted separation. The same value was obtained applying the statistical methodology of \citet{deruiter77}} \label{maxdistance_numobj}
\end{figure}

In order to build a sample of X--ray emitters with optical counterparts, we performed cross--matching 
of the X--ray and optical catalogs
using the \textsc{topcat} tool \citep{taylor05},  
applying a criterion of minimum sky distance along a great circle. X--ray sources located inside the field covered by optical data
and their optical counterparts are represented in Figure \ref{xray_opt_counterparts}. \\
We chose a maximum search radius of 2\,arcsec after carrying out several tests considering 
a maximum searching distance of 1 to 5\,arcsec with a bin of 0.5\,arcsec (Figure \ref{maxdistance_numobj}). 
As expected, the number of counterparts increases with the maximum search radius, but also the 
number of multiple matches within the error circle. Setting a search radius of 1\,arcsec we minimise 
the number of multiple cross--matches, but at the cost of loosing an outstanding number of potential counterparts. 
On the other hand, choosing a search radius greater than 2\,arcsec produces a large number of multiple matches, 
increasing the uncertainty in counterpart identification. Therefore, we set a 2\,arsec 
search radius as an adequate compromise. This practical criterion is consistent with the more rigorous analysis 
performed below.

We detected 340 optical counterparts of the X--ray sources in at least one optical band. 
Of them, 322 are detected in the B band, 332 in V,  
334 in R and 317 in the I band. 

We carried out a quantitative analysis of the completeness\footnote[1]{Fraction of all real associations 
between the X--ray and optical sources that are indeed classified as identifications.} and 
reliability\footnote[2]{Fraction of claimed identifications expected to be true counterparts.} of our cross--matching 
procedure by 
applying the methodology of \citet{deruiter77}. 
The separation between sources is defined by the dimensionless variable $r$ as: 
\begin{center}
\begin{equation}
{r = (\frac{\Delta\alpha^{2}}{\sigma_{\alpha}^{2}} + \frac{\Delta\delta^{2}}{\sigma_{\delta}^{2}})^{1/2},}
\end{equation} 
\end{center}

\noindent where $\Delta\alpha$ and $\Delta\delta$ are the position differences between the X--ray source and its 
optical counterpart in RA \& DEC, and $\sigma_{\alpha}^{2}$ and $\sigma_{\delta}^{2}$ are the overall astrometric errors 
computed as $\sigma=(\sigma_{X}^2 + \sigma_{opt}^2)^{1/2}$. In this work, we assume $\sigma_{X}$\,=\,0.7\,arcsec and $\sigma_{opt}$\,=\,0.3\,arcsec. $\sigma_{X}$ was calculated by square adding a systematic uncertainity of 0.5 and a centroiding error of 0.5 arcsec.

\noindent The likehood ratio is defined as:

\begin{center}
 \begin{equation}
LR(r) = \frac{dp(r|id)}{dp(r|c)} = \frac{1}{2\lambda} e^{\frac{r^{2}}{2} (2\lambda - 1)}
\end{equation}
\end{center}

\noindent where:
\begin{itemize}
 \item dp(r$|$id) is the \textit{a priori} probability that an X--ray source and its optical counterpart have a separation 
 between $r$, and $r + dr$ due to astrometrical errors
 \item dp(r$|$c) is the probability that a confusing background optical object is found in the range $r$, and $r + dr$ 
 from the X--ray source position.
 \item  $\lambda$ measures the number of confusing objects within the error circle and depends on the optical 
 objects density \textbf{($\rho_{opt}$)} as: $\lambda = \pi\sigma_{\alpha}\sigma_{\delta}\rho_{opt}$.
\end{itemize}

An optical source can be considered as a true counterpart of an X--ray source if its LR is higher than a certain 
threshold value L. L is obtained maximising the sum of completeness and reliability that
are, in turn, mathematically defined in terms of the probabilities defined above. In our case,
L\,=\,0.56 and its corresponding variable is $r$\,=\,2.6 which gives us a separation distance of 2.0 arcsec, a value consistent with the adopted one. 
The completeness of our detection is 98.3\% (i.e. some 1 object lost due to the use of a particular cutoff value L), 
while reliability is 87.3\% ($\sim$ 39 potentially false matchings).

\section{Analysis}
\label{sec_analysis}

\subsection{X--ray properties}\label{sec:xray_properties}

\begin{figure}[!ht]
\centering
\includegraphics[width=7.5cm,angle=0]{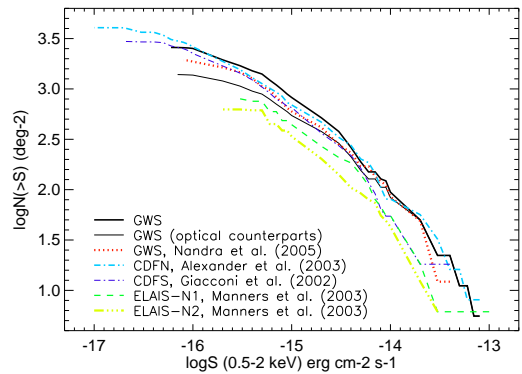}
\protect\caption[ ]{Cumulative logN-logS functions for the Groth field in the soft band, for all detected X--ray sources 
(thick solid line) and for X--ray sources with optical counterpart (thin solid line). 
Five distributions, with different exposure times and effective areas, have also been 
represented as comparison: GWS by \citet{nandra05} (dotted line), CDFN (thick dash-dot-dash line), CDFS (thin dash-dot-dash line), 
ELAIS-N1 (dash line) and ELAIS-N2 (dash-dot-dot-dash line)\label{logN_logS}}
\end{figure}

We have computed the cumulative number count distributions (logN-logS) per deg$^{2}$ in the soft band 
and compared it with other surveys in order check the reliability of our source detections. We have chosen 
the soft band to be able to 
compare our distribution with that computed by \citet{nandra05} and those provided by other deep 
X--ray surveys, Chandra Deep Field South \citep[CDFS,][]{giacconi02}, Chandra Deep Field North  
\citep[CDFN,][]{alexander03}, ELAIS-N1 and ELAIS-N2 \citep{manners03}. All distributions are represented 
in Figure \ref{logN_logS}. For our sample two curves have been computed, one for the complete sample of 
X--ray detections and the other only for the X--ray detections that have optical counterparts. 
The median flux levels are 1.81\,$\times$\,10$^{-15}$ and 1.97\,$\times$\,10$^{-15}$ \,erg\,cm$^{-2}$\,s$^{-1}$, 
respectively. For the objects brighter than $\sim$\,10$^{-14}$\,erg\,s$^{-1}$\,cm$^{-2}$ both distributions 
coincide, while for objects fainter than $\sim$\,10$^{-14}$\,erg\,s$^{-1}$\,cm$^{-2}$ the density of X--ray sources with optical counterparts starts to decrease,
since the optical sample is incomplete at the depth of our X--ray data. In general, at our flux limit the 
number counts of our complete X--ray sample are in good agreement with Nandra et al. (2005) and other 
surveys. We have detected a higher density of sources with flux below $\sim$\,5\,$\times$\,10$^{-15}$\,erg\,cm$^{-2}$\,s$^{-1}$, most likely due to the use of different detection 
algorithms and more sensitive detection threshold. As we already mentioned, our threshold 
is 2\,$\times$\,10$^{-7}$, two times higher than the one chosen by Nandra et al. (2005) 
(see \ref{xray_detection} for more information). \\

\begin{figure}[!ht]
\centering
\includegraphics[width=7.5cm,angle=0]{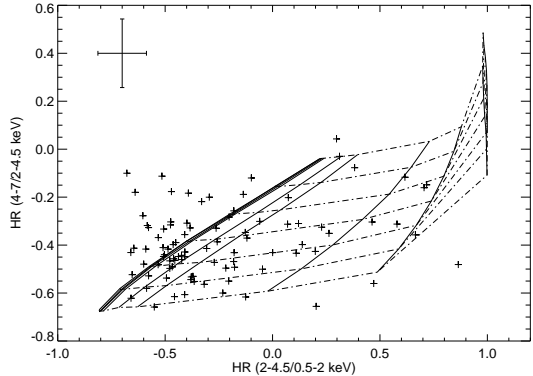}
\protect\caption[ ]{X--ray color--color diagram based on 2--4.5/0.5--2keV (HR$_1'$) vs. 4--7/2--4.5keV (HR$_2'$) hardness ratios. X--ray sources (crosses), from the complete X--ray catalog, are represented and superposed on an absorbed power law model grid. Photon index is varying between 0 and 3 in steps of
0.5 (dashed lines from up to down) and a log(N$_{H}$) between 20 and 24 in steps of 0.5 (solid lines from left to right). Median error bars are represented in the upper left corner of the diagram.
\label{hr_grids}}
\end{figure}

Three independent energy bands, soft, hard2 and vhard (see the Section \ref{xray_observations} for ranges), 
have been used to obtain the X--ray color--color diagram shown in Figure \ref{hr_grids}. Hardness ratios have been 
calculated as described in Section \ref{xray_detection}. X--ray sources are superimposed on a model grid computed using 
power law Spectral Energy Distributions (SEDs) with different photon indexes, $\Gamma$, affected by a galactic absorption column density, N$_{H}$, of 
1.3\,$\times$\,10$^{20}$\,cm$^{-2}$ and different values of intrinsic absorption; the \chandra\  
ACIS-I response matrices have been also included in the computation. To obtain the fluxes we have used XSPEC 
\citep{arnaud96}, varying the photon index between 0 and 3 with a step size of 0.5, and the intrinsic 
absorption column density between 20 and 24 (in logarithmic scale) in steps of 0.5. 
The majority of our sources are located within the model grid region, suggesting that their X--ray SED 
is compatible with the assumed model. A specially high concentration of objects can be seen in the region of N$_{H}$ between 10$^{20}$ and 5\,$\times$\,10$^{21}$\,cm$^{-2}$ and with $\Gamma$\,$\simeq$\,1.8. A group of sources are located outside the model grid, showing a soft excess, suggesting that they can not be properly described by a single power law. For the fitting of these objects probably more components should be added. 

\subsection{Optical structural parameters}
\label{optical_structural}

We obtained a set of structural parameters of the optical counterparts of our X--ray sample, 
in order to describe quantitatively their structure and morphology. In this section we describe 
the procedure used to derive the parameters and to produce a morphological classification. 

We used SExtractor v2.2.2 \citep{bertin96} to detect and extract the sources. 
Details of the source extraction process are provided in \citet{cepa08}. 
The version used (v2.2.2) was modified by B. Holwerda
to measure the Abraham concentration index \citep{abraham94}. 
This is defined as the ratio between the 
integrated flux within certain radius defined by the normalized radius $\alpha$ and the total flux. We used the values obtained at $\alpha$\,=\,0.3 in our analysis in order to compare our results 
with those of \citet{abraham96}.
Another relevant output parameters for our study are the CLASS\_STAR, 
which permits to roughly classify objects as compact or extended 
(for more information see Section \ref{morph_class}), and 
the background that will be used later.  

\begin{figure}[ht!]
\centering
\includegraphics[width=7.5cm,angle=0]{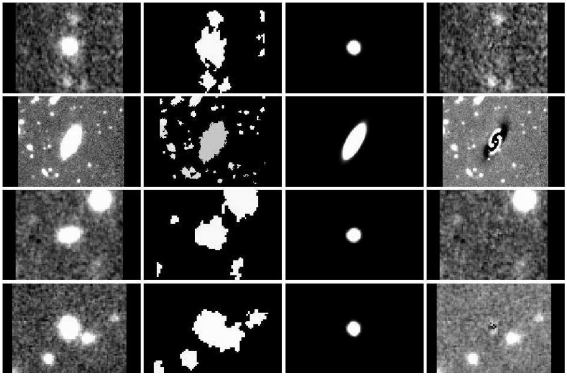}
\protect\caption[ ]{Four sample images of the GIM2D decomposition procedure. First column shows subimages 
(obtained from the scientific image) of objects to be fitted; the second column represents the mask subimages, 
third column shows the subimages obtained from the modelling and the fourth column shows the residual images, 
obtained from the original image after subtracting the model 
\label{im_imseg}}
\end{figure}

\begin{figure}[ht!]
\centering
\includegraphics[width=7.5cm,angle=0]{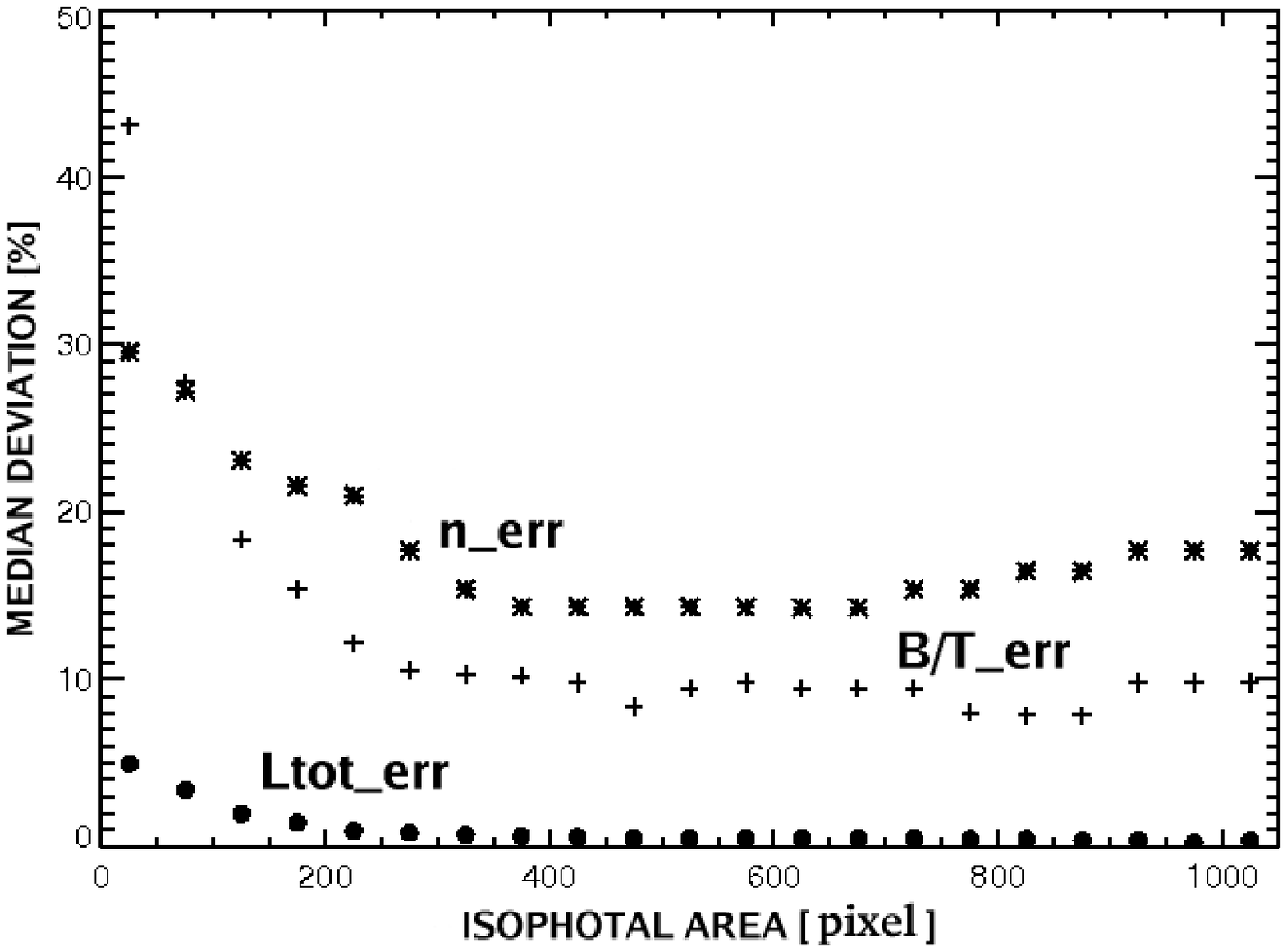}
\caption[ ]{Evolution of the median error of the bulge-to-total ratio B/T (crosses), S\'ersic index \textit{n} (stars) and
total flux (filled circles)  L$_{\textit{tot}}$ as functions of the isophotal area. Deviations are evaluated as explained in Section \ref{optical_structural}.
\label{fig_dev}}
\end{figure}

In addittion, we used the Galaxy IMage 2D software package \citep[GIM2D,][]{simard98,simard02} 
versions 2.2.1 and 3.1, to perform the galaxy structural parameter decomposition (see Figure \ref{im_imseg}).  
This software considers two components: an spheroid, represented by a 
S\'ersic profile\footnote{While any rigourous analysis of the structural components 
of an AGN host should include an additional component, namely a pointlike nucleus, 
the small size of most of our objects compared with the seeing disk prevents from 
implementing such a detailed approach. Rather, we have simply assumed that the sum of 
the nuclear and bulge components can be roughly represented by a single S\'ersic profile.}, 
and  a disk represented by an exponential profile.
Using the standard steps in DAOPHOT \citep{stetson87} we obtained 
the PSF images required for GIM2D execution. The initial parameters (including background estimation) and setup for GIM2D were 
determined using the images and catalogs produced by SExtractor \citep{bertin96}.
We obtained from the output provided by GIM2D the following 
parameters\footnote{According to the available documentation \citep{simard98,simard02}, 
GIM2D calculates the Abraham concentration at four normalized radii 
$\alpha$\,=\,0.1, 0.2, 0.3 and 0.4. \citep{abraham94,abraham96}. Comparing the results with the
ones obtained with SExtractor, a systematic shift of $\sim$\,0.2 is found towards larger values
in GIM2D figures. After performing several tests we concluded that the GIM2D code
performs its computations using normalized \textit{area} rather than 
normalized \textit{radius} against what is stated in the documentation.}: 
bulge-to-total ratio (B/T), residual parameter (\textit{R}) 
(Schade et al. 1995, 1996),  asymmetry (A) index \citep{abraham94,abraham96}, 
the S\'ersic index (n) and the total luminosity ($L_{tot}$). 
The residual parameter and the asymmetry index quantify the galaxy irregularity, by measuring
the deviation from the simple, symmetric model (e.g. spheroid + disk), using as input the residual 
and original image, respectively. 
GIM2D also calculates for certain output
parameters their lower and upper limits at 99\% 
confidence level. In order to estimate the accuracy of the best fit parameters 
we have looked at the deviation, computed as the maximum between the best value
and the upper or lower limits (see Figure \ref{fig_dev}). 
In this respect, the B/T parameter  and S\'ersic index have revealed 
inaccurate  when applied to  faint and small objects ($\lesssim$ 300 pixel, while
the seeing disk (FWHM) area is about 30 pixels) as shown in Figure \ref{fig_dev}. 
Therefore we have decided to do not use these parameters for the morphological
classification. 

In practice, we observed that the code either fails or produces unreliable output parameters in 
a number of cases: (1) when the source R magnitude is less than 18 or larger than 24, (2) when the 
isophotal area is very small (usually less then 90 pixels), (3) when the 
central object is surrounded by close neighbouring sources or nearby bright companions, and (4) when 
the source is close to the frame boundaries.

\subsubsection{Morphological classification}
\label{morph_class}

The selection of compact objects (QSOs, faint host AGNs or active stars) 
has been accomplished by using the 
SExtractor CLASS\_STAR output parameter. The computed value depends on the seeing, the peak
intensity of the source and its isophotal area. The parameter ranges between 0 (extended
object like a galaxy) and 1 (pointlike object).
A source has been deemed as compact if its CLASS\_STAR parameter is larger than  0.9.

To determine the morphology of extended objects 
(i.e. not selected as compact according to the criterion above), we analyzed four 
parameters: Abraham concentration index (C), Abraham asymmetry 
index (A), B/T ratio and residual parameter (\textit{R}). We carried out the analysis in three
different bands (B, V, and R) in order to determine the most reliable parameters, and the best band for morphological classification. We excluded I band here due to fringing.

\begin{figure}[ht!]
\centering
\includegraphics[width=7.5cm,angle=0]{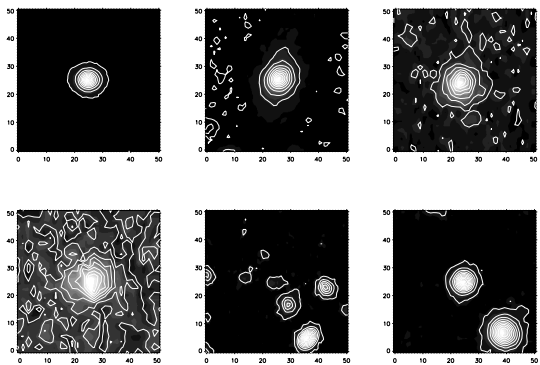}
\protect\caption[ ]{Sample of different Hubble type galaxies and their contour diagrams; Up: Left: Group I  (E/S0),
Center: group II (Early type spiral),  Right:  group III (Spiral); Down: Left, group III  (Spiral), Center: group IV (Irr) , Right:  compact object (C\,$\geq$\,0.7 or CLASS\_STAR\,$>$\,0.9). All objects are located in the center of the images.
\label{perfiles}} 
\end{figure}

We first classified each object using the set of
parameters mentioned above, separately for each parameter in 
each band, considering the results from previous works as 
limits between different morphological types. We used 
the results of \citet{im02} and \citet{simard98} to 
obtain preliminary morphology with the B/T parameter, \citet{schade95}
 results for the residual parameter and \citet{abraham96}
 for the classification with the concentration and 
asymmetry index. 
In practice, we found that, in many cases, different parameter
sets lead to a different classification of a given object even in the same band. 
In order to have an independent evaluation criterion, we also carried out a visual classification of the
objects. Despite its subjectiveness, this procedure can yield highest--quality results 
when applied to bright and extended objects, thus being a convenient method to 
test the reliability of the structural parameters that will be in turn 
applied to the analysis of fainter objects. The visual classification has been
accomplished by inspection of the radial profile, 2D surface plot and 
isophotal countours diagram of each source (see Figure \ref{perfiles}). For the dimmest and smallest objects of 
our sample (galaxies with the magnitude R\,$>$\,24 and isophotal area $<$ 100 pixels) the profiles and isophotal contours are too 
noisy to get an acceptable result by means of a simple visual inspection.

From this comparison we conclude that for our sample, characterised by faint
objects (typically R $\gtrsim$ 22) with small isophotal area ($\lesssim$ 300 pixel, while
the seeing FWHM disk area is about 30 pixels) 
the most reliable parameter is the concentration index C, combined either with 
the assymmetry index or the residual parameter. We therefore based our morphological classification in the concentration index
combined with both asymmetry index and residual parameter.
The R band has been used, since, on the one hand, the largest part of optical counterparts have been found in this band, 
and on the other, the achieved S/N is generally larger. 
B, V, and I bands have  been used to perform the morphological classification of objects not detected
in the R band, or detected in this band but with very low S/N ratio.

Based on the classification using visual inspection of the brightest sample objects, and upon comparison with previous 
works \citep[e.g.][see references above]{abraham96} in the present paper we created five categories or groups 
associating ranges of values of C with Hubble types:

\begin{center}
{\mbox{0: C\,$\geq$\,0.7 or CLASS\_STAR\,$\geq$\,0.9\,$\rightarrow$\,Compact obj.}\\
I: 0.45\,$\leq$\,C\,$<$\,0.7\,$\rightarrow$\,E, E/S0 and S0\\
II: 0.3\,$\leq$\,C\,$<$\,0.45\,$\rightarrow$\,S0/S0a-Sa\\
III: 0.15\,$\leq$\,C\,$<$\,0.3\,$\rightarrow$\,Sab-Scd\\
IV: C\,$<$\,0.15\,$\rightarrow$\,Sdm-Irr\\}
\end{center}

The first group is populated by elliptical
and lenticular galaxies, whose profiles do not show any
trace of disk or spiral arms (see Figure \ref{perfiles}),
the second group includes very early spiral type galaxies which
show signs of spiral arms. The third group comprises spiral galaxies 
while  the fourth includes late type spirals and
irregular
galaxies, that show almost no regular form in
their profiles. All galaxies, even those in
the first group (elliptical and lenticular) have a
concentration index C\,$\leq$\,0.7, while all
objects with C\,$>$\,0.7 have been classified by SExtractor as
compact ones, with the CLASS\_STAR parameter $>$\,0.9.

\begin{figure}[ht!]
\centering
\includegraphics[width=7.5cm,angle=0]{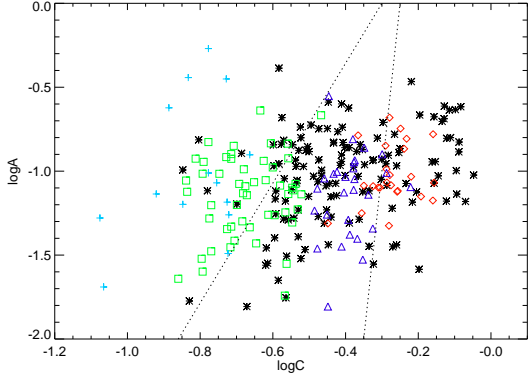}
\protect\caption[ ]{Morphological classification diagram 
combining the Abraham asymmetry index (A) vs. Abraham 
concentration index (C). Diamonds present E and S0 galaxies, 
triangles S0/S0a-Sa galaxies, squares young spiral 
galaxies (Sab-Sbcd) and crosses late type galaxies (Sdm-Irr). Compact objects
are represented by star symbols. Dotted lines separate the locii of elliptical, spiral and peculiar galaxies (from right to left) obtained by \cite{abraham96}. There is a small shift of our concentration and asymmetry index towards smaller values due to the influence of the atmospheric seeing conditions.
\label{plots_morph}}
\end{figure}

In Figure \ref{plots_morph} we depict the Abraham
asymmetry vs. concentration index diagram segregated according to the morphology
determined using the visual classification
for the brightest objects, and the structural parameters
(in most of the cases C combined with A) for the fainter
ones. Comparing our Figure \ref{plots_morph} with that shown in \citet{abraham96}, 
it can be seen that in our case there is a small shift
of the concentration index towards smaller values.
This can be explained considering the influence of
the atmospheric seeing conditions. SExtractor does not apply any correction
on the computed parameters for this effect, and as the seeing induced PSF increases,
both concentration and asymmetry indexes tend to decrease, as confirmed by our
simulations carried out by convolving simple galaxy models with seeing PSFs of different
widths. Since  the \citet{abraham96} diagram has been
obtained from HST data, these differences are expected. 
\\  

\begin{figure}[ht!]
\centering
\includegraphics[width=7.5cm,angle=0]{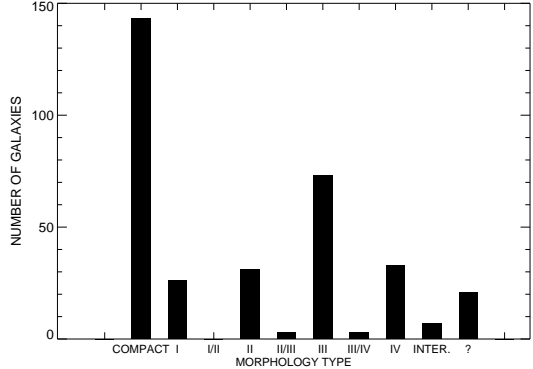}
\caption[ ]{The histogram represents the final morphological classification of the GWS X--ray emitters with optical counterparts. Different values of the Abraham concentration index correspond to different groups I-IV (see text). A group of possible 
interacting objects has been detected; a group of sources (marked with '?') remained unidentified due to their very low 
S/N ratio.
\label{morph_hist}}

\end{figure}

The morphological analysis described above has been applied to the complete sample of 340 X--ray objects 
with optical counterpart (see Figure \ref{morph_hist}). Of those, 333 were classified using the R band,
4 sources using the V band and finally 3 objects by means of B band data. 
The largest group (143 sources) is composed by  
compact objects (QSOs, faint host AGNs, 
stars) followed by group III (73 galaxies). 57 objects are found either in group II or III, having outstanding 
bulge component a signs of a formed disk and spiral arms.
Only 33 galaxies are found to belong to the IV group. Finally, we detected 7 closed pairs, whose possible physical relation will be confirmed when photometric redshifts are computed (paper III, in preparation).
Due to the high noise level, or to the presence of other objects close to the observed ones, the
classification uncertainty for some of the objects is quite
high, and it was not possible to determine the
morphological group: 3 galaxies could be either in group II or III and other 3 either
in group III or IV. For 21 sources it was not possible
to determine the morphology since they are at the detection limit. 

\section{Relationship between X--ray and optical structural properties}
\label{sec_rel}

\subsection{Nuclear classification}\label{sec:nuclear_class}

The simplest way to classify the nuclear type is based on the X--ray-to-optical ratio (X/O). The typical value of X/O flux ratios for X--ray selected AGN (both type 1 or BLAGN and type 2 or NLAGN) 
is in the range
between 0.1 and 10 \citep[e.g.][]{fiore03}. For optically selected type 1 AGNs the typical value is X/O\,$\simeq$\,1
\citep{alexander01,fiore03}.
At high X/O flux ratios
(well above 10) we can find BLAGN and NLAGN as well as high-z high-luminosity obscured AGN (type 2 QSOs), 
high-z clusters of galaxies and extreme BL Lac objects. Finally, the X/O\,$<$\,0.1 region is typically populated by 
coronal emitting stars, normal galaxies (both
early type and star-forming) and nearby heavily absorbed (Compton thick) AGN.

\begin{figure*}[ht!]
\centering
\includegraphics[width=8cm,angle=0]{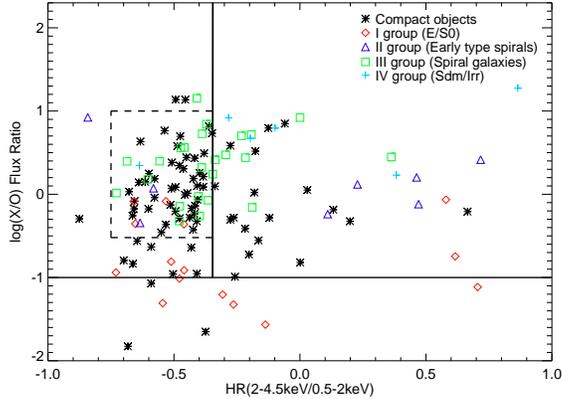}
\protect\caption[ ]{Relationship between X/O ratio and HR$(2-4.5keV/0.5-2keV)$, for the 
different morphological types. Solid lines separate BLAGN and NLAGN regions and the area with X/O\,$<$\,0.1, 
where stars, normal galaxies and Compton thick AGNs can be found \citep{fiore03}. 
Dashed line box presents the limits obtained by \citet{della04} where $\sim$\,85\% of their spectroscopically
 identified BLAGN have been found. In our case, inside of this box can be found $\sim$\,51\% of all objects from the 
 diagram.        
\label{XO_HR-h2s}}
\end{figure*}

We have applied a simple criterion for performing a coarse nuclear--type classification of
our sample objects, based on diagnostic diagrams relating X--ray-to-optical ratios (X/O) to
hardness ratios (HR). This criterion is based on the results of \citet{della04} on the \xmm\
Bright Serendipitous Survey; they found that, when plotting the HR$(2-4.5keV/0.5-2keV)$
(equivalent to our HR$_1'$) against X/O, computed as the ratio of the observed
X--ray flux in the 0.5-4.5keV (soft+hard2) energy range and the optical R-band flux, most (85\%) of the
BLAGN identified by means of optical spectroscopy are tightly packed in a small
rectangular region of the diagram, while NLAGN (type 2) tend to populate a wide area of the diagram
(towards harder values of the hardness ratio).
We applied the same approach to our sample, combining the HR$_1'$ hardness ratio with
the X/O ratio, where the optical flux $F_R$ has
been derived from the Petrosian magnitudes in the R band. The resulting diagram, that also includes host
morphology information, is shown in Figure
\ref{XO_HR-h2s}, where the dashed--line box indicates the locus of the BLAGN region. We set the right
edge of this box, HR$(2-4.5keV/0.5-2keV)$\,$\approx$\,-\,0.35 as a coarse boundary separating 'unobscured'
AGN (BLAGN) and 'obscured' AGN (NLAGN). On the other hand, we consider the small box depicted in the diagram as the ``highest probability'' region for finding a BLAGN \citep{della04}. We find that a large fraction of our objects,
63\% \footnote{All fractions given here are relative to
the total number of X--ray objects having optical
counterpart and detected in both soft and hard2 energy bands such as HR$(2-4.5keV/0.5-2keV)$ can be computed}, fall inside the region of BLAGNs, while 51\% of the sample
is placed within the small BLAGN box
defined by \citet{della04}.
Regarding morphology, we do not find evidence of any relationship
between nuclear and morphological types, since we did not obtain
any clear separation of nuclear species based on morphological type. Nevertheless, we can see that the
majority of objects identified as compact (65\%) are placed in
the BLAGN region. It is also interesting to highlight the relationship
between the X/O ratio and the host galaxy morphology, with early-type
galaxies having generally lower values for X/O ratio and an opposite behavior of late-type objects.
Finally, we find that only about 7\% of our sample objects are placed in the lower region of the diagram
(i.e. that typical of stars/normal galaxies/Compton-thick AGN).

We also represented the X/O ratio with respect to HR$_1$, yielding a basically identical distribution as that
described above, and with respect to HR$_2$ and HR$_2'$, observing that in these two cases is harder to get a 
clear separation between different nuclear/morphological types; this behavior is expected since hardness ratios 
computed from the hard and vhard bands are less sensitive to absorption than those involving the soft band, which is one 
of the main criteria for the separation between 
broad and narrow line AGN.

\subsection{X--ray and optical properties}

\begin{figure*}[ht!]
\centering
\begin{minipage}[c]{.49\textwidth}
\includegraphics[width=8cm,angle=0]{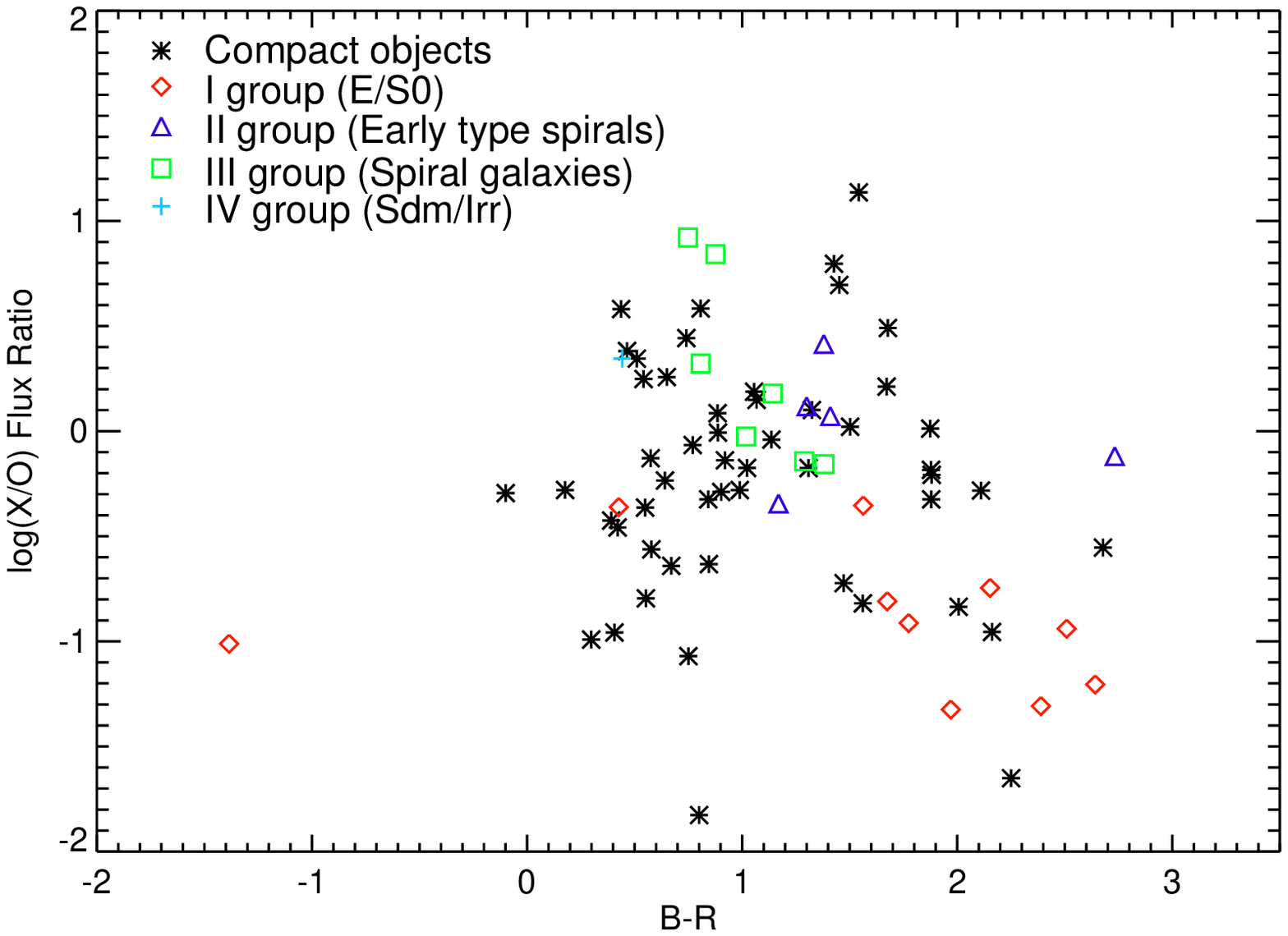}
\centering
\end{minipage}
\begin{minipage}[c]{.49\textwidth}
\centering
\includegraphics[width=8cm,angle=0]{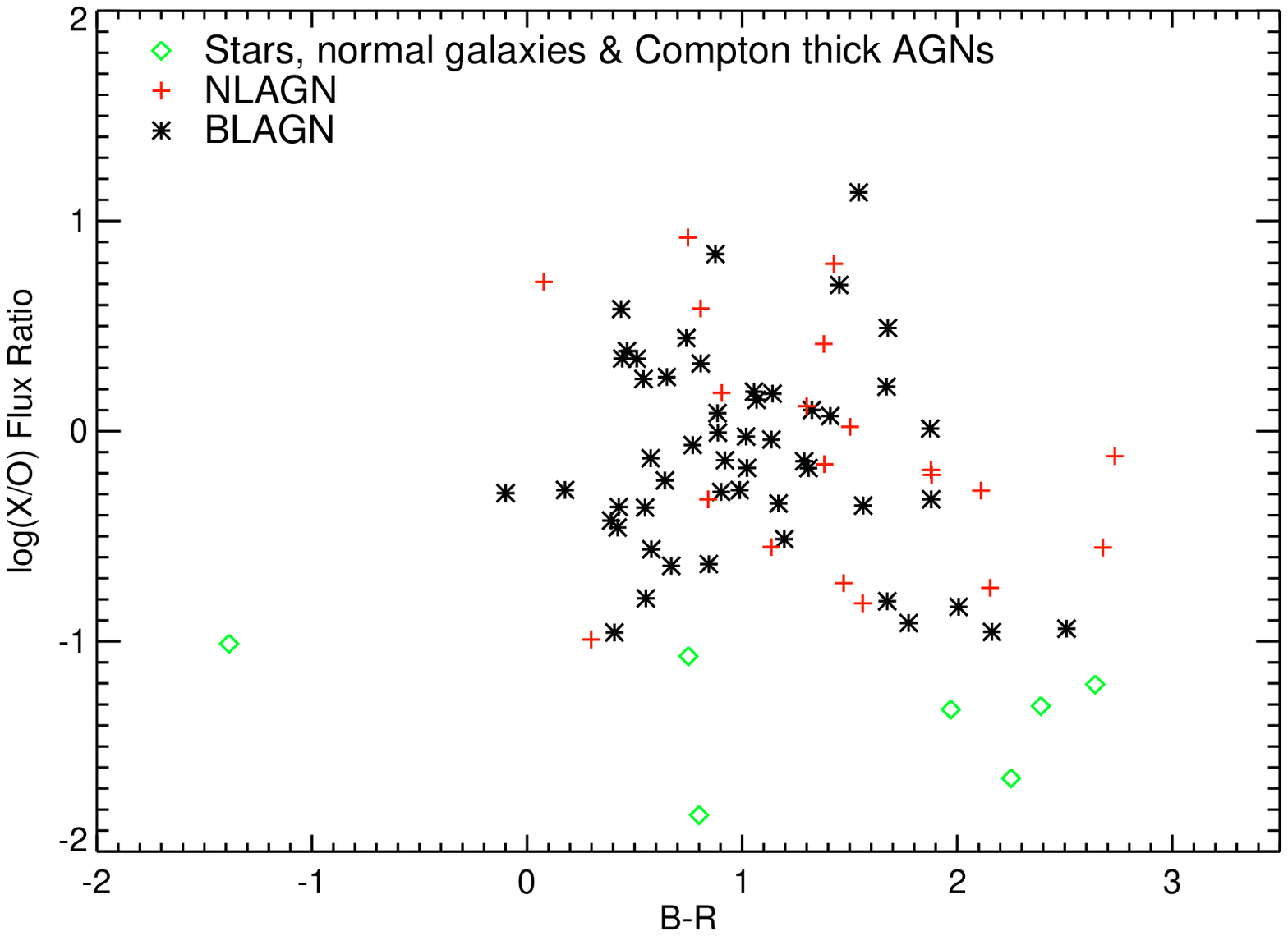}
\end{minipage}
\protect\caption[ ]{Log(X/O) flux ratio vs. B-R color for different morphology (left) and nuclear types (right).
\label{xo_color}}
\end{figure*}

We combined the X/O ratio with optical colors (B-R and B-I) for the
different morphological groups and nuclear types, as shown in Figure \ref{xo_color}.
We observe that compact and late-type objects tend to show bluer colors, while
early-type galaxies tend to present redder colors and also lower X/O values, as expected. On the other hand, there is a mild
tendency for BLAGN to show bluer colors than those found in NLAGN, as already 
observed
in local universe samples \citep[e.g.][]{yee83,sanchez04}. 
The B--R average value for BLAGNs is 1.0
and for NLAGNs 1.4,  with $\sigma$\,=\,0.5 and 0.6 respectively. Using the Kolmogorov-Smirnov test, the two distributions are
statistically different with a significance level of 99\%.

\begin{figure*}[ht!]
\centering
\begin{minipage}[c]{.49\textwidth}
\includegraphics[width=8cm,angle=0]{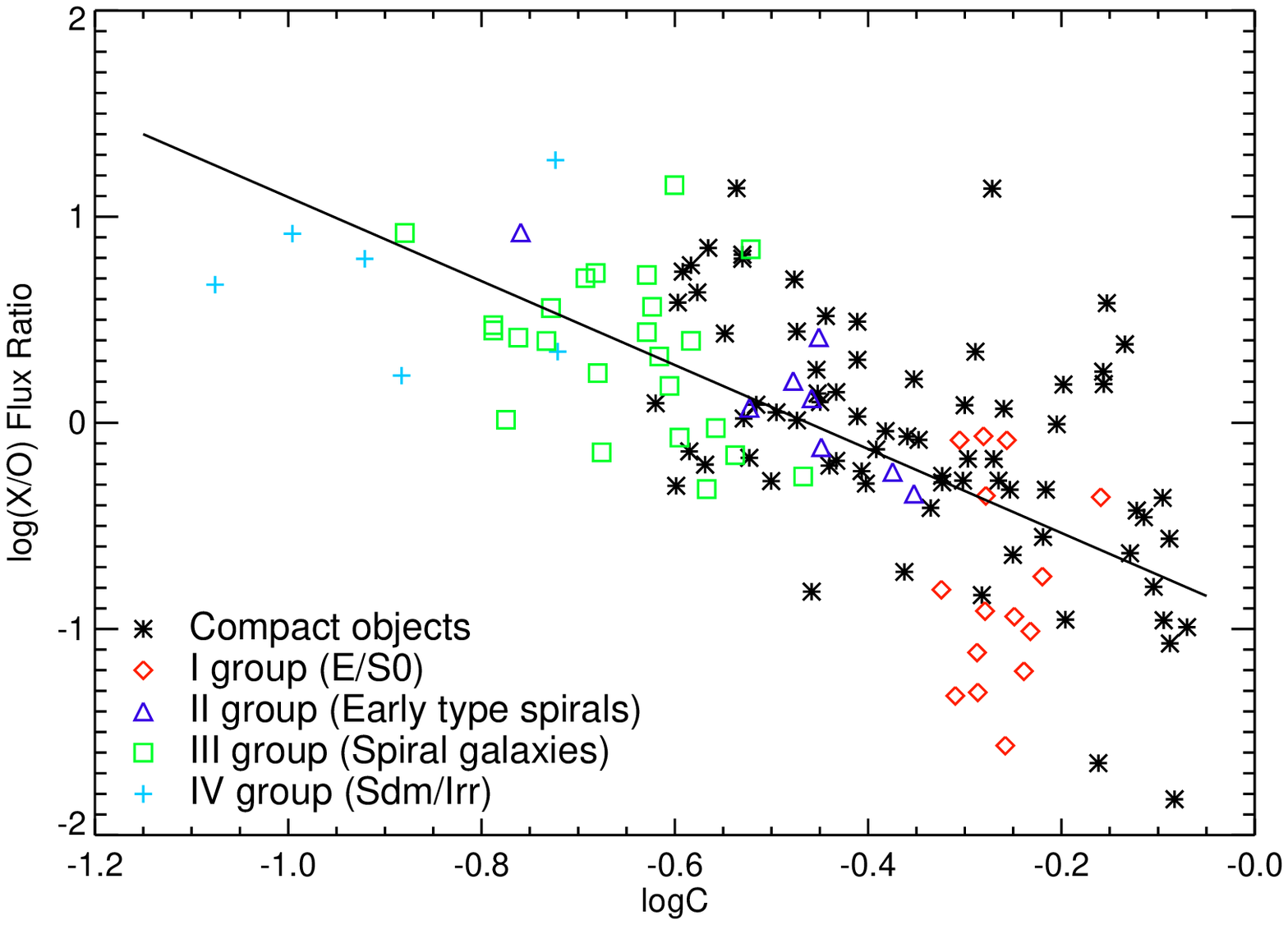}
\centering
\end{minipage}
\begin{minipage}[c]{.49\textwidth}
\centering
\includegraphics[width=8cm,angle=0]{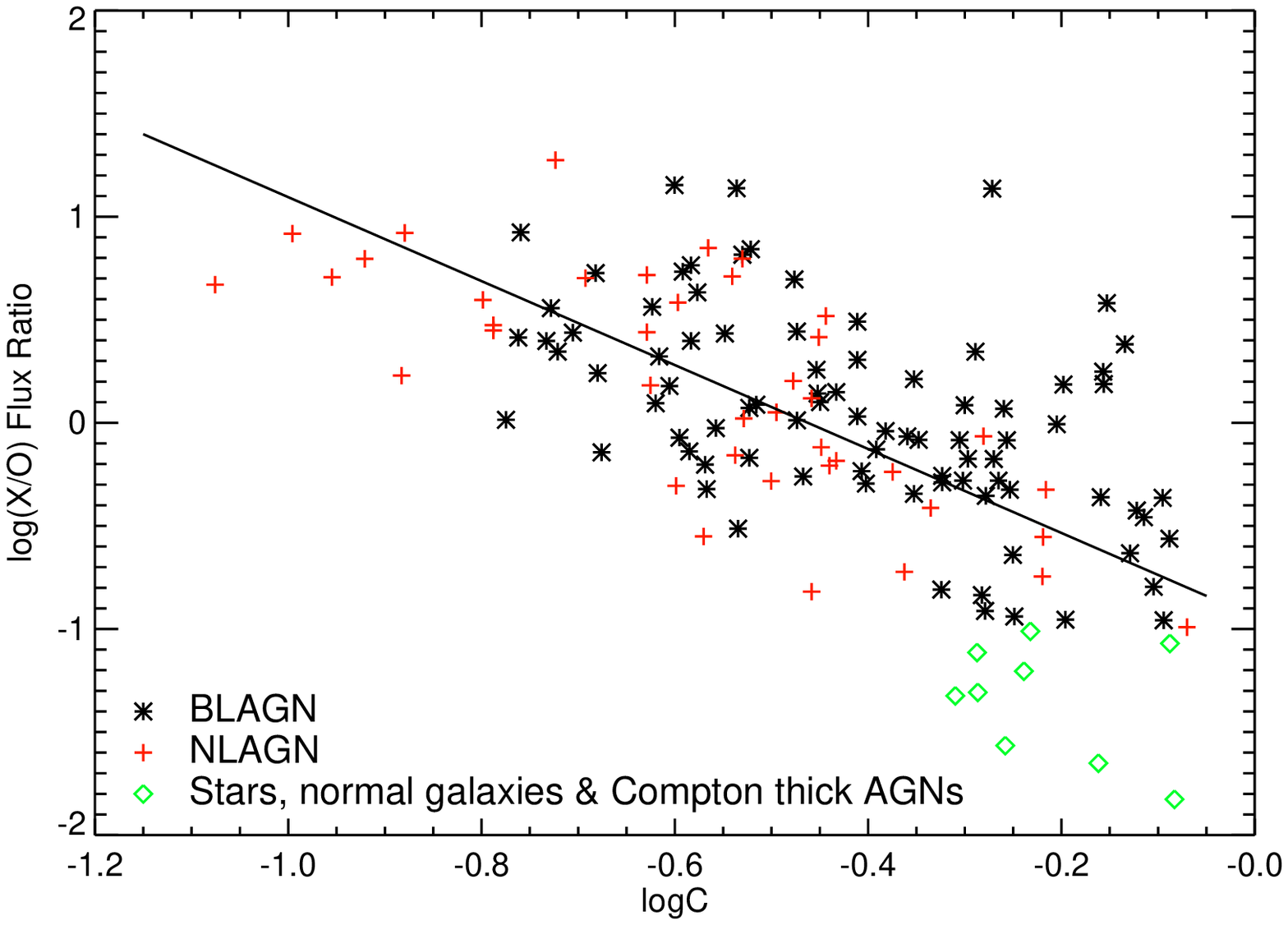}
\end{minipage}
\protect\caption[ ]{Relationship between X/O and the Abraham concentration index C for different morphological (left) 
and nuclear types (right). A clear anticorrelation is observed. 
\label{XO_C}}
\end{figure*}

\begin{figure*}[ht!]
\centering
\includegraphics[width=8cm,angle=0]{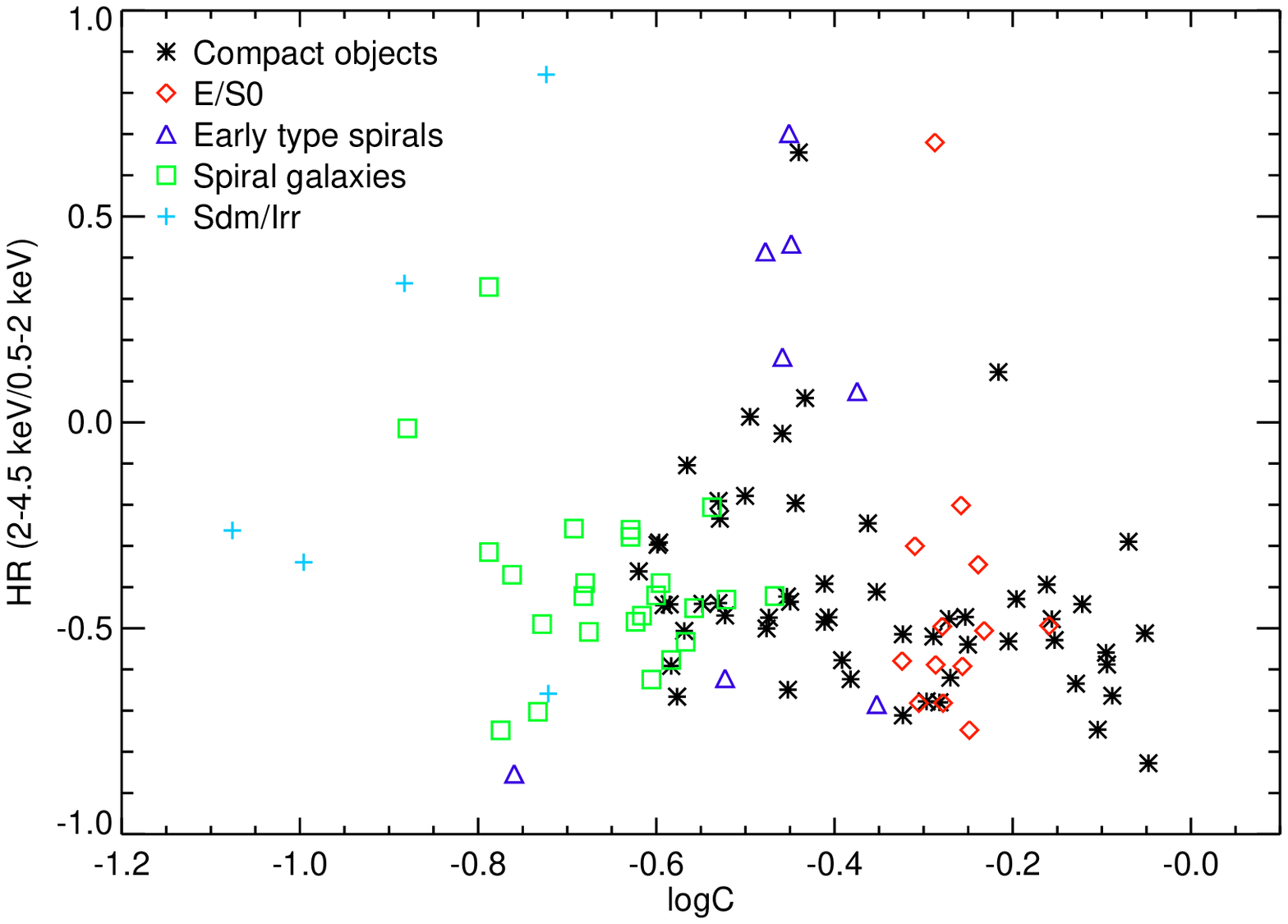}
\protect\caption[ ]{Relationship between the 
hardness ratio HR$(2-4.5keV/0.5-2keV)$, and the concentration index C for the 
different morphological types.
\label{HR_logC}}
\end{figure*}

When representing the X/O ratio vs. the Abraham concentration index C we find a strong
anticorrelation between these two parameters (Figure \ref{XO_C}), 
obtaining a 
correlation coefficient of  $\sim$\,-0.5. Using Spearman and Kendall (S--K) statistics,
we obtained a correlation significance level $>$99\%. The first order polynomial 
function has been fitted with a slope of -2.03,  $\log$\,X/O-intercept of -0.94 
and a standard deviation of $\sim$\,0.5. This result is consistent 
with that referred above and depicted in Figure~\ref{XO_HR-h2s}, that is the tendency of
early-type galaxies to have lower than average X/O values and the opposite behavior in
late-type ones. As seen in Figure~\ref{XO_C}, all morphological groups, including compact objects, tend to follow the same relation. For compact objects we obtained similar parameters using S--K statistics: a correlation coefficient of $\sim$\,0.5 and a correlation significance level $>$99\%. A least squares fit gives to the anticorrelation a slope of -2.4 and a $\log$\,X/O-intercept of -0.95. Comparing different nuclear types, we found that both BLAGN and NLAGN follow the same distribution 
(see Figure~\ref{XO_C}). On the other hand, 
we did not find any clear correlation between HR$(2-4.5keV/0.5-2keV)$ and the concentration index (Figure~\ref{HR_logC}), suggesting that the observed anticorrelation is not related to obscuration. 
\\

The physical origin of the observed anticorrelation is not well established. First of all, it is not easy to disentangle possible bias effects that superimposed to variable seeing conditions. It has been pointed out the existence of redshift-dependent biases in both A and C, due to the reduction of apparent size and surface brightness with respect to the sky background with increasing z, along with poorer sampling and lower S/N \citep{Brinchmann98, Conselice03}; on the other hand, there is a ``bandpass shift'' with respect to the rest wavelength \citep{Bershady00}. Even if these effects combine to produce some uncertainty in the morphological classification of galaxies, generally in  the sense of shifting objects to apparently later Hubble types (i.e. with lower C values), A and C seems to be reproducible out to z\,=\,3, within a reasonable scatter \citep{Conselice03}. Moreover, there can be a truly evolutionary effect in the AGN host galaxies, indicating that galaxies at higher redshift are intrinsically less concentrated \citep{grogin05}. 

At this stage we cannot apply any correction which depends on the redshift, although this is one of the objectives for the future work. However, we have performed a preliminary evaluation by representing the concentration index C versus the spectroscopic redshifts z gathered from the DEEP2 catalog for 84 objects from our sample. While for non compact objects we obtained an anticorrelation between these two parameters, C and z are found uncorrelated for compact objects, in clear contrast with the observed behavior in Figure~\ref{XO_C}.  

So, although the biases described above might contribute partially to it, the observed anticorrelation seems to also be a result of a true physical effect. In order to attempt to explain it, we have first to understand the physical connections of the two observables C and X/O. The concentration index (C) is tightly 
linearly correlated to the central velocity dispersion \citep{graham01a,graham01}, which
is in turn related to the nuclear black hole (BH) mass \citep[e.g.][]{magorrian98}. Therefore,
the concentration and the nuclear BH mass (M$_{BH}$) are correlated. The physical connections of the X--ray to optical flux ratio are not so evident. The 
X/O ratio measures the X--ray flux (in the 0.5-4.5keV band), normalized with the R-band optical flux of 
the whole galaxy (nucleus, bulge and disk). It should be noted that no bandpass, or K-correction is performed; 
therefore if the sample spans a large
redshift range the interpretation of the quantity becomes more problematic. If the nuclear luminosity is large compared with that of the host galaxy, X/O ratio can be thought as a 
measure of the X--ray to optical 
spectral index, 
$\alpha_{OX}$\footnote{This relation is usually defined as $\alpha_{OX} = 0.3838 \log (f_{2keV}/f_{2500\AA})$ }. 
This parameter has been found to be strongly anticorrelated with the UV luminosity L$_{2500\AA}$, without a significant
 correlation with redshift \citep{steffen06}. On the other hand \cite{bian05} found a strong correlation between the 
 hard/soft X--ray spectral index, $\alpha_X$ and the Eddington ratio in a sample of 41 BLAGN and narrow-line Seyfert 1 
 (NLS1) galaxies observed with ASCA. Other works have confirmed this correlation \citep{grupe04, shemmer06}.
 The relationship between $\alpha_X$ and $\alpha_{OX}$ has not been thoroughly studied in
large samples of AGN, but a linear correlation has been found between both spectral indexes in
a sample comprising 22 out of 23 quasars in the complete the Palomar-Green X--ray sample with z\,$<$\,0.4 and M$_B$\,$<$\,-23 \citep{shang07}. If this last correlation holds for our sample, X/O could be tracing the Eddington ratio in the large nuclear luminosity limit. In addition, if the host galaxy luminosity is large when compared with the nuclear luminosity ($L_{bulge+disk} \gg L_{nucleus}$),  
the X/O ratio  can be thought as a lower bound of the X--ray-to-bulge luminosity ratio, that is in turn a (weak) measure of the
AGN Eddington ratio $L/L_{Edd} \propto L/ M_{BH}$,  assuming that the X--ray luminosity represents the nuclear luminosity
and the bulge luminosity is proportional to the bulge mass (and therefore to  M$_{BH}$).  

 Hence, we can guess a (loose) correlation between X/O and the energy production efficiency
of the AGN. Under these assumptions, the C vs. 
X/O relation traces the correlation between the Eddington ratio and the nuclear BH mass. 
The obtained result could therefore suggest that early-type galaxies, having poor matter 
supply to feed the
activity, have lower Eddington rates than those of late-type, with larger reserves of the gas for AGN feeding. 

This suggested approach is consistent with the results obtained by \citet{ballo06,ballo07} 
in a sample of X--ray selected AGNs at 
z\,$\leq$\,1. 
 They observed that a large fraction  of low-luminosity 
AGNs are fed by black holes with a mass 
 M$_{BH}$\,$>$\,3\,$\times$\,10$^{6}$\,M$\odot$, 
and Eddington rate $\ll$ 1, while 
a small ( $<$\,10\%) fraction of AGNs host small SMBH (M$_{BH}$\,$<$\,10$^{6}$\,M$\odot$)
and Eddington rate $\approx$ 1, which corresponds to a more efficient accretion rate. 
They stated that their results strongly suggest that most of the low-luminosity X--ray
selected AGNs (at least up to z\,$\leq$\,0.8) are powered by rather massive black holes, 
experiencing a low-level accretion in a gas-poor environment.

Our approach is also consistent with the results  of \citet{wu04}, who studied the
black hole masses and Eddington rates of a  
sample of 135 double--peaked broad line AGNs, observed in two 
surveys: SDSS and a survey of BLAGN radio emitters. 
They obtained that if the separation between the line peaks (which is correlated to the SMBH mass) 
decrease, the Eddington rate increases. 

\cite{kawakatu07} found an anticorrelation between the mass of a SMBH and the luminosity 
ratio of infrared to active galactic nuclei Eddington luminosity, L$_{IR}$/L$_{Edd}$, for a sample of ultraluminous 
infrared galaxies with type 1 Seyfert nuclei (type 1 ULIRGs) and nearby QSOs, which also could be consistent with our 
approach. The anticorrelation is interpreted as a link between the mass of a SMBH and the rate of mass 
accretion onto a SMBH, normalized by the AGN Eddington rate, which indicates that the growth of the black hole is 
determinated by the external mass supply process, and not the AGN Eddington-limited mechanism, changing its mass 
accretion rate from super-Eddington to sub-Eddington.

\section{The clustering of X-Ray emitters and extended optical counterparts}\label{clustering}

The study of large-scale environmental conditions of AGN host galaxies are useful
not only for understanding the differential properties of the AGN phenomena, but to trace the
formation and evolution of such galaxies and the structures where they reside. Despite the
relevance of X--ray selected AGNs, their clustering properties still remain quite unknown
\citep[][and references therein]{basilakos05}. In this sense,
and taking into account the characteristics of the X--ray source catalog presented in this
work, we have calculated the 2-point angular correlation function (2p-ACF) for significant
subsamples.

\begin{figure}[!h]
\centering
\includegraphics[width=8cm,angle=0]{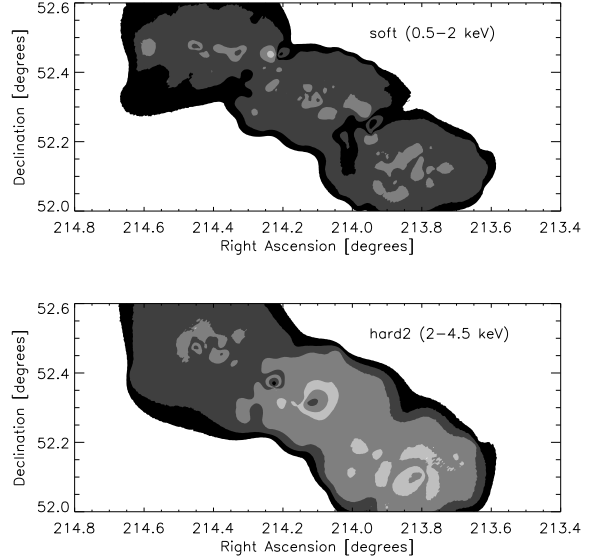}
\protect\caption[ ]{Smoothed sensitivity maps in the soft (0.5--2 keV) and hard2 (2--4.5 keV) bands, used
to convolve the uniform density random catalogs. The sensitivity dimension is normalized to the maximum
exposure time in each band. Contours are linearly binned and lighter regions correspond to higher
sensitivities.
\label{sensitivity}}
\end{figure}

The 2p-ACF, $\omega(\theta)$, gives the excess of probability, with respect to a random homogeneous
distribution, of finding two sources in the solid angles $d\Omega_1$ and $d\Omega_2$ separated
by an angle $\theta$, and it is defined as
\begin{equation}
dP={\cal{N}}^2\ [1+\omega(\theta)]\ d\Omega_1 d\Omega_2
\end{equation}
where $\cal{N}$ is the mean number density of cataloged sources. To measure
the 2p-ACF we have follow the same procedure as done in \citet{cepa08}, adopting the estimator
of \citet{landyszalay93}, which can be written in the form
\begin{equation}
\omega(\theta_i)={DD-2DR+RR\over RR}
\end{equation}
where $DD=N_{ss}(\theta_i)$ is the fraction of possible {\it source-source} pairs counted in
$i$-bins over the angular range studied and $DR=\left [ {(N_g-1)\over 2N_r}\right ] N_{gr}(\theta_i)$,
$RR=\left [ {N_s(N_s-1)\over N_r(N_r-1)}\right ] N_{rr}(\theta_i)$ are the normalized counts of {\it source-random}
and {\it random-random} pairs in these bins, respectively. To make the random catalogs
we placed $(200\times N_{source})$ random
points by following the X--ray source angular distribution. Howewer, the source density is affected by several
instrumental biases such as detector gaps and sensivity variations due to survey regions with
different exposure times. To account for these effects, we made normalized sensitivity maps from the effective
exposure time and source distributions for each band used in this section (soft and hard2). The raw maps were
modelled using a mimimum-curvature algorithm. Next, the uniform density
random catalogs were convolved with the corresponding sensitivity
map to obtain differential number counts in the surveyed area. In the case of the
HR$(2-4.5keV/0.5-2keV)$ catalog, we adopt the sensitivity map used for the hard2 source one. The
smoothed footprint of each sensitivity map can be seen in Figure~\ref{sensitivity}.

We measure $\omega(\theta_i)$ in $i-$logarithmic
bins of width $\Delta(\log\ \theta)=0.2$ in the angular scales $0.24\leq\theta\leq24.0$ arcmin. The lower
limit in angular separation is according to the minimum detection scale defined in the Section~\ref{xray_detection}. The formal
error associated with the 2p-ACF measured is the Poissonian estimate
$\sigma^2_{\omega}=\frac{1+\omega(\theta_i)}{DD}$.
The 2p-ACF binned measures were fitted using
the $A_{\omega}(\theta^{1-\gamma} - C)$ power law. We adopted
a fixed canonical slope $\gamma$=1.8 (the so called ``comoving clustering model'') and C is defined as
\begin{equation}
C=\frac{\sum\ N_{rr}(\theta)\ \theta^{-0.8}}{\sum\ N_{rr}(\theta)}
\end{equation}
corresponding to the numerical estimation of the ``integral constraint'' \citep{peebles80}. As in
\citet{miyaji07}, we use the normalization $A$ as the fitting parameter rather than the correlation length
$\theta_c=A^{1/(\gamma -1)}$, since the former gives better convergence of the fit.

From the whole catalog we have selected the soft (0.5--2 keV) and hard2 (2--4.5 keV) sources (given
their similar depths in the survey),
the X--ray objects with HR$(2-4.5keV/0.5-2keV)$\,$\neq$\,0 and the fraction of X--ray sources with optical
counterparts. Table~\ref{table2} summarizes the basic results of the correlation analysis for these
subsamples. In all cases, we have obtained significant positive clustering signal. Regarding the soft band
sample, the correlation length $\theta_c$ derived from $A_{\omega}$ is consistent with the obtained
by \citet{basilakos05} (XMM-{\it Newton} LSS) and it is $\sim$\,2\,$\sigma$ larger than the values obtained
by \citet{gandhi06}, \citet{carrera07} and \citet{ueda08}. In the specific case of the hard2 band, we only
found in the literature the 2p-ACF estimations of \citet{miyaji07}, valid for X--ray sources detected
in the XMM-{\it Newton} observations of the COSMOS field and corresponding to their MED band: our
fitted correlation length is $\sim$\,4.3 times larger, although our sample is a factor 2 deeper.

\begin{table*}[!ht]
\begin{center}
\caption{Clustering properties of the X--ray selected sources. $N_{source}$ is the number of data points
used in the 2p-ACF estimation and $A_{\omega}$ is the best-fit value in the separation range given
by $\theta_{\rm min}-\theta_{\rm max}$, assuming a power-law
$\omega(\theta)=A_{\omega}\ \theta^{(1-\gamma)=-0.8}$. The goodness-of-fit statistics ($\chi^2$) and
the fitted 2p-ACF at 1 arcmin [$\omega(1')$] are given in the final columns. The errors represent
1$\sigma$ uncertainties.}
\bigskip
\label{table2}
\begin{tabular}{cccccc}
\hline
\hline
{Subsample}&$N_{source}$&$\theta_{\rm min}-\theta_{\rm max}$&$A_{\omega}\ (1-\gamma=-0.8)$&$\chi^2$&$\omega(1')$\\
&&arcmin&$\times10^{-3}\ {\rm degrees}$&&\\
\hline
soft (0.5-2 keV) & 465 & 0.24-24.0 & $6.99\pm0.68$ & 0.030 & $0.15\pm0.02$ \\
hard2 (2-4.5 keV) & 266 & 0.24-24.0 & $9.19\pm1.28$ & 0.107 & $0.20\pm0.03$ \\
HR$_{(2-4.5keV/0.5-2keV)}$ & 190 & 0.24-24.0  & $9.68\pm1.30$ & 0.109 & $0.21\pm0.03$ \\
Optical counterparts (I$\leq$23) & 104 & 0.60-24.0 & $19.99\pm3.32$ & 0.096 & $0.42\pm0.07$ \\
\hline
\end{tabular}
\end{center}
\end{table*}

\begin{figure}[!h]
\centering
\includegraphics[width=8cm,angle=0]{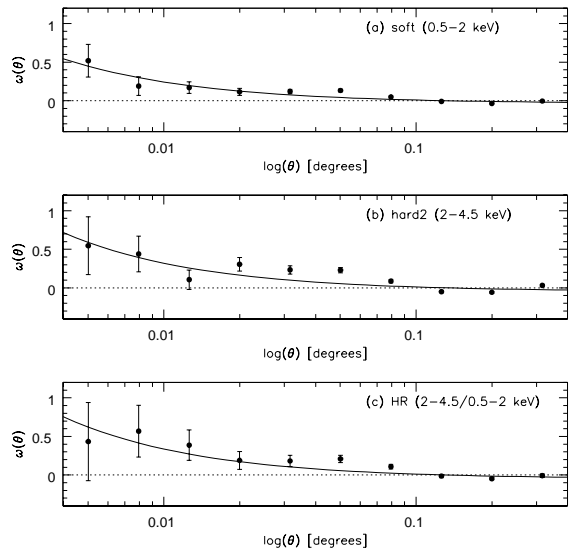}
\protect\caption[ ]{2p-ACF estimates of the X--ray sources detected in the soft
band (panel a), hard2 band (panel b) and HR as defined in eq. 1 (panel c). The
continuous line represents the $A_{\omega}\ \theta^{(1-\gamma)=-0.8}$ power law fit to the data.
\label{soft_hard_hr_cf}}
\end{figure}

In Figure~\ref{soft_hard_hr_cf} are represented our 2p-ACF measurements with the best-fit
$A_{\omega}\ \theta^{-0.8}$ power-law in soft and hard2 bands (panels a and b, respectively). It is
noticeable that the clustering amplitude obtained in both bands looks similar. Thus, in the case of
the Groth field, we would not expect redshift or spatial clustering differential effects in the soft
and hard2 band samples. Interestingly, an absence of correlation in differently defined hard bands
has been reported by \cite{gandhi06} and \citet{puccetti06}; nevertheless, \citet{yang03} and \citet{basilakos04}
not only yield a positive correlation amplitude, but the hard band signal is several times that of the soft
band. Other results with similar correlation lengths in both bands are given by \citet{carrera07},
\citet{miyaji07} and \citet{yang06}.
\\

In the panel c of Figure~\ref{soft_hard_hr_cf} is represented the 2p-ACF estimations fo a HR$(2-4.5keV/0.5-2keV)$
selected sample. The clustering amplitude is slightly larger than the soft or hard2 band ones. Moreover,
the whole HR$(2-4.5keV/0.5-2keV)$ subsample was decomposed using the boundary referred in Sect.~\ref{sec:nuclear_class}
to separate type 1 (or BLAGN) from type 2 (or NLAGN) populations and a positive, very similar clustering signal
from both subsamples emerged from the 2p-ACF estimations, despite their goodness-of-fit statistics are
worse than the tabulated values in a factor $\sim$\,1 because of the small size of the subsamples.

Finally, we measured the 2p-ACF of the optical counterparts of X--ray sources in order to compare
the clustering behavior with those that come from the analysis of galaxy populations in the same field.
For a reliable comparison is necessary to diskard the compact objects (i.e. with CLASS\_STAR$>$0.9)
from the optical counterparts catalog, as well as all objects fainter than a limiting magnitude. We
have selected the peak of the apparent brightness distribution in the I-band as such limiting magnitude.
Thus, X--ray counterparts with I$>$23 were ruled out, remaining 104 extended objects, which can be regarded
as AGN host candidates of the corresponding X--ray sources.

In spite of this skinny sample, it is possible to obtain some interesting lines of thought about the X--ray
selected AGN clustering. It is suggested that the typical environments of AGNs are no different to
those of inactive galaxies in general, at redshift ranges above z$\sim$0.4 \citep[e.g.][]{grogin03,waskett05}.
Nevertheless, \citet{coil04} and \citet{gilli05} have argued that the AGN are preferentially hosted
by early-type galaxies at a typical redshift of $\sim$\,1. In this way, \citet{basilakos04}
found that clustering length values of hard X--ray objects are comparable with those of Extremely Red
Objects (EROs) and luminous radio sources. More recently, \citet{geor07}
demonstrated that X--ray-selected AGNs at z\,$\sim$\,1, in the AEGIS field, avoids underdense regions at
99.89\% of significance.

\begin{figure}[ht!]
\centering
\includegraphics[width=8cm,angle=0]{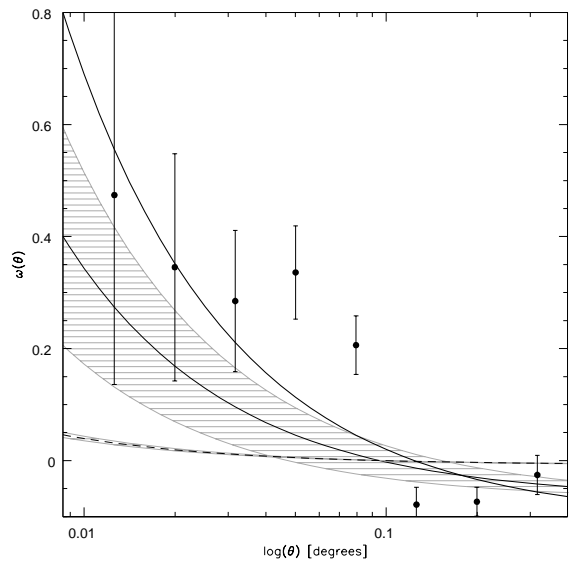}\protect\caption[ ]{
2p-ACF estimation for the optical extended (CLASS\_STAR$\leq$0.9)
counterparts with I magnitude brighter than 23 (solid circles). The
continuous thick line represents the $A_{\omega}\ \theta^{(1-\gamma)=-0.8}$ power law fit to the data.
The dashed line represents the 2p-ACF estimation for galaxies with I$\leq$23
in the Groth field, whereas the continous thin line it does for the V-I$>$3 color selected galaxies
with I\,$\leq$\,24, both extracted from \citet{cepa08}. The shaded regions represent fitting errors of
1$\sigma$.
\label{counter_cf}}
\end{figure}

In Figure~\ref{counter_cf} we have depicted the 2p-ACF of the optical counterparts selected as
described above. The 2p-ACFs of the Groth field galaxy sample with I$\leq$23 only and with V-I$>$3
(I$\leq$24) from \citet{cepa08} are overplotted. At least in the range 0.01$\leq \theta \leq$0.1 degrees, the
clustering signal produced by the optical counterparts of our X--ray sources is 1-2$\sigma$ stronger than
the corresponding to the I$\leq$\,23 galaxy population. In terms of fitted correlation length, the $\theta_c$(X--ray)
(1-$\gamma$=-0.8) is $\sim$\,30 times larger than that of the Groth field galaxies, but only $\sim$\,2.3 times
larger than the maximum correlation length of the V-I$>$3 galaxies in the same field. In the other hand, using the optical
data from the counterparts catalog, we found that the median V-I observed color is 1.86 and
most of 65\% of the optical counterparts have an observed color V-I$>$2. This means that a significant
population of our optical counterparts is a composition of red and very red galaxies, whereas its clustering signal tends
to reproduce that associated with such galaxies, which drive the $\sim$10x clustering excess found by
\citet{cepa08}. At once, more than a 55\% of the optical counterparts belong to the morphological categories I and II,
as defined in the Section~\ref{morph_class}.

\section{Summary and Conclusions}
\label{sum_conc}

We processed and analyzed public, 200 ksec \textit{Chandra}/ACIS
observations of three fields comprising the original GWS gathered from the 
\textit{Chandra} Data Archive, combined with our broadband survey carried 
out with the 4.2m WHT at La Palma. Our X--ray catalog contains 639 unique 
X--ray emitters, with a limit flux at 3$\sigma$ level of 4.8\,
$\times$\,10$^{-16}$\,erg\,cm$^{-2}$\,s$^{-1}$ and a median flux of 1.81\
$\times$\,10$^{-15}$\,erg\,cm$^{-2}$\,s$^{-1}$ in the full (0.5-7 keV) band. 
Cross-matching the X--ray catalog with our optical broadband catalog, we have 
found 340 X--ray emitters with optical counterparts. The completeness and the 
reliability of cross-matching procedure are 97\,\% and 88.9\,\%, 
respectively. At our flux limit, the number counts of our complete X--ray 
sample are in good agreement with Nandra et al. (2005) and other comparison 
surveys mentioned in Section~\ref{sec:xray_properties}. X--ray SEDs of the mayority 
of our sources are compatible with the power law model, affected with the 
galactic absorption column density and the intrinsic absorption. We have performed a
morphological classification obtaining and analyzing different structural parameters, 
including the bulge-to-total ratio, residual parameter, S\'ersic index, 
asymmetry and concentration indexes. We also performed the classification 
based on visual inspection. To obtain the nuclear classification, 
we applied a criterium based on the correlation between X/O 
flux ratio and HR1' hardness ratio.
63\,\% of our sources, with HR1'\,$\ne$\,0, have been classified as BLAGNs. 
The electronic table summarizes the X/O flux ratios, HR1' values and morphological 
classification of all 340 X--ray emitters with optical counterparts.
We have investigated the 
correlations between X--ray and broadband optical structural 
parameters that provide with information about the host galaxy morphology 
and populations, finding that:

1. For a sample where most of the objects are faint (R\,$\ge$\,22) and with 
small isophotal area ($\le$\,300 pixels), bulge-to-total flux ratio has 
revealed inaccurate to perform the morphological classification. 
Concentration index combined with the asymmetry index have been found 
to provide the best separation criterion for such a sample. 
different morphological types have been observed to distribute according to the following 
ranges of concentration index: group 0 (compact objects) $\rightarrow$  C\,$\geq$\,0.7 
(or CLASS\_STAR\,$\geq$\,0.9), 
group I (E, E/S0 and S0) $\rightarrow$ 0.45\,$\leq$\,C\,$<$\,0.7, 
group II (S0/S0a-Sa) $\rightarrow$  0.3\,$\leq$\,C\,$\leq$\,0.45 (0.47), 
group III (Sab-Scd) $\rightarrow$ 0.15\,$\leq$\,C\,$\leq$\,0.3 and group IV 
(Sdm-Irr) $\rightarrow$ C\,$\leq$\,0.15  

2. No clear separation has been found between the morphology and the nuclear 
type, but most of our objects classified as compact are placed in the region 
of BLAGNs. 

3. We can confirm the tendency of compact and late type objects to show bluer
 colors, while early type galaxies tend to present redder colors. On the other hand, we 
 find a mild tendency of BLAGNs to show bluer colors when compared with NLAGNs. 

4. We obtained an anticorrelation between the X-to-optical flux ratio
 and the concentration index (which parametrizes the morphology), with significance level higher 
 than 99\,$\%$, showing that the early-type galaxies tend to have lower and 
 the late-type galaxies higher than average X/O values. Objects classified 
 as compact show this anticorrelation as well, with the same significance level. 
 The anticorrelation might suggest that early-type galaxies, having poor matter 
 supply to feed the activity, have lower accretion rates than those of late type, 
 with larger reserves of gas for AGN feeding.

Finally, we presented the first results on the two-point angular correlation
function of the X--ray selected sources in the Groth field from \textit{Chandra}/ACIS
observations, segregated in four subsamples: soft, hard2, HR$_{(2-4.5keV/0.5-2keV)}$
and optical extended counterparts. We obtain the following results and conclusions:

1. All subsamples reveal a significant positive clustering signal in the
$\sim$\,0.5 to 6 arcmin separation regime. Canonical power-law ($\gamma$=1.8) fits to
the angular correlation functions have been made, obtaining correlation lengths
between $\sim$\,7 and 27 arcsec, depending on the subsample.

2. We have neither found substantial differences between the clustering of soft and hard2
selected X--ray sources, nor when comparing the populations of type 1 and type 2 AGN
(i.e. 'unobscured' and 'obscured' AGN, respectively), separated using the
HR$(2-4.5keV/0.5-2keV)$\,$\approx$\,-\,0.35 boundary.

3. The clustering analysis of the optical counterparts of our X--ray emitters provides a
correlation length significantly larger than the one found for the whole galaxy population with
the same limiting magnitude, and same field,
but similar to that of strongly clustered populations like red and very red (bulge-dominated)
galaxies. A clustering signal $>1\sigma$ was observed even at $\theta \sim30$ arcsec, which
corresponds to 160 $h^{-1}$ kpc, assuming a mean redshift of 0.85 for \textit{Chandra}
extragalactic sources \citep{hasan07}. This comoving distance is, for instance, comparable to
the separation of two mergers in the t$\approx$2 Gyr stage of their large-scale evolution, according
to the model of \citet{mayer07} on rapid formation of SMBH. Hence, our results suggest that
the environment plays an important (but not unique) role as possible triggering mechanism of
AGN phenomena. This fact can be perfectly compatible with the interpretation given to the
anticorrelation between X/O ratio and the concentration index. The former may be associated to
the genesis of the AGNs, while the anticorrelation could be that to their maintenance and evolution.

\section{Acknowledgements}

This work was supported by the Spanish {\it Plan Nacional de
Astronom\'\i a y Astrof\'\i sica} under grants AYA2005--04149 and AYA2006-2358.
JIGS and JGM acknowledge financial support from the Spanish Ministerio 
de Educaci\'on y Ciencia under grants AYA2005-00055 and  AYA2006-2358, respectively. 

This research has made use of software provided by the Chandra X--ray Center (CXC) 
in the application packages CIAO and ChIPS.

IRAF is distributed by the National Optical Astronomy Observatory, which is operated
by the Association of Universities for Research in Astronomy (AURA) under cooperative
agreement with the National Science Foundation.

This publication makes use of data products from the Two Micron All Sky Survey, which is
a joint project of the University of Massachusetts and the Infrared Processing and Analysis
Center California Institute of Technology, funded by the National Aeronautics and Space
Administration and the National Science Foundation.

We would like to thank Laura Tom\'as who kindly provided us with her software for diagnosis 
of hardness ratios using model grids.

We acknowledge support from the Faculty of the European Space Astronomy Centre (ESAC).

\bibliographystyle{aa}

\end{document}